\documentclass[11pt]{article}
\usepackage{authblk}
\usepackage[left=0.75in,right=0.75in]{geometry}
\usepackage{empheq}
\usepackage{amssymb}
\usepackage{dsfont}
\usepackage{slashed}
\usepackage[pdfstartview=FitH,pdfpagemode=None]{hyperref}
\usepackage{color}
\usepackage{multirow}
\usepackage{mathtools}
\usepackage{stackrel}
\numberwithin{equation}{section}

\newcommand{\nn}{\nonumber}
\newcommand{\be}{\begin{equation}}
\newcommand{\ee}{\end{equation}}
\newcommand{\ba}{\begin{eqnarray}}
\newcommand{\ea}{\end{eqnarray}}
\newcommand{\bea}{\begin{eqnarray}}
\newcommand{\eea}{\end{eqnarray}}

\begin{document}
\title{Functional Determinants of Radial Operators in $AdS_2$}
\author[1]{Jerem\'ias Aguilera-Damia\thanks{jeremiasadlp@gmail.com}}
\author[2]{Alberto Faraggi\thanks{alberto.faraggi@unab.cl}}
\author[3,4]{Leopoldo Pando Zayas\thanks{lpandoz@umich.edu}}
\author[3]{Vimal Rathee\thanks{vimalr@umich.edu}}
\author[1,4]{Guillermo A. Silva\thanks{silva@fisica.unlp.edu.ar}}
\affil[1]{Instituto de F\'isica de La Plata - CONICET \& Departamento de F\'isica - UNLP  C.C. 67,

 1900 La Plata, Argentina}
\affil[2]{Departamento de Ciencias F\'isicas, Facultad de Ciencias Exactas, Universidad Andr\'es Bello, Sazie 2212, Piso 7, Santiago, Chile}
\affil[3]{Leinweber Center for Theoretical Physics, University of Michigan, Ann Arbor, MI 48109, USA}
\affil[4]{The Abdus Salam International Centre for Theoretical Physics, Strada Costiera 1,

34014 Trieste, Italy}
\date{}
\maketitle
\begin{abstract}
We study the zeta-function regularization of functional determinants of Laplace and Dirac-type operators in two-dimensional Euclidean $AdS_2$ space. More specifically, we consider the ratio of determinants between an operator in the presence of background fields with circular symmetry and the free operator in which the background fields are absent. By Fourier-transforming the angular dependence, one obtains an infinite number of one-dimensional radial operators, the determinants of which are easy to compute. The summation over modes is then treated with care so as to guarantee that the result coincides with the two-dimensional zeta-function formalism. The method relies on some well-known techniques to compute functional determinants using contour integrals and the construction of the Jost function from scattering theory. Our work generalizes some known results in flat space. The extension to conformal $AdS_2$ geometries is also considered. We provide two examples, one bosonic and one fermionic, borrowed from the spectrum of fluctuations of the holographic $\frac{1}{4}$-BPS latitude Wilson loop.
\end{abstract}

\newpage
\tableofcontents
\section{Introduction}

The AdS/CFT correspondence is, in its simplest form, a strong/weak duality  where weakly couple gravity is equivalent to strongly coupled field theory  \cite{Aharony:1999ti}.  For some time much of the effort was focused on this window.  Exploring and understanding the full power of the AdS/CFT correspondence requires the need to go decisively beyond the leading order agreement. Going beyond the leading order is a time-honored tradition in physics; it suffices to recall that the leading energy levels in the hydrogen atom can be obtained using the Bohr model which lacks a  proper description of the relevant degrees of freedom.  Supersymmetric localization techniques in field theory have now provided predictions for the gravity results beyond the leading answer \cite{Pestun:2007rz} setting the stage for systematic explorations beyond the leading order.

For general observables, semiclassical physics is our only systematic approach to probe the AdS/CFT correspondence beyond the leading classical limit.  The main precept of semi-classical physics consists in integrating quadratic quantum fluctuations around a well-defined classical background.  When we get down to practical evaluations, however,  we must face the sometimes messy process of treating divergences, as typical of quantum field theory but now with the added intricacies of being in curved space-time.  Determining the semiclassical one-loop effective action is equivalent, by definition,  to the computation of determinants.

 There are many situations in the AdS/CFT correspondence where one ends up comptuting determinants in $AdS_2$ and its generalizations. The original  discussion of the holographic dual to the $\frac{1}{2}$-BPS Wilson loop made used of $AdS_2$  determinants for the first time \cite{Drukker:2000ep}.  The list of one-loop effective action problems that can be tackled exploiting the fact that $AdS_2$ is a homogeneous space is rather large. For example, it naturally includes the one-loop effective actions of supersymmetric D3 and D5 branes dual to Wilson loops in ${\cal N}=4$ SYM in the symmetric and anti-symmetric representations, respectively \cite{Gomis:2006sb,Gomis:2006im}. Given that the worldvolume of these configurations are $AdS_2\times S^2$ and $AdS_2\times S^4$, the one-loop effective actions reduce also to determinants on $AdS_2$  \cite{Faraggi:2011bb,Faraggi:2011ge,Buchbinder:2014nia}. A similar class of one-loop effective action appears also in the context of ABJM \cite{Muck:2016hda}. In the context of localization of supersymmetric field theories there have been some natural appearances of $AdS_2$ \cite{David:2016onq,Cabo-Bizet:2017jsl,Rodriguez-Gomez:2017kxf,David:2018pex}. Determinants of $AdS_2$ operators have also figured prominently in logarithmic corrections to the entropy of extremal black holes \cite{Sen:2011ba}.  When the worldvolume geometry is not $AdS_2$ new methods need to be developed; we have recently discussed in fair detail the case of the  $\frac{1}{4}$-BPS holographic Wilson loop \cite{Aguilera-Damia:2018twq} using the results of the present paper.  
  
Motivated by the above richness of applications, in this manuscript we discuss determinants of  general Laplace and Dirac operators in asymptotically $AdS_2$ spacetimes.  We use the regularization method chosen {\it par excellence} in curved spaces: $\zeta$-function regularization.  These methods have a long an fruitful history, dating back over four decades, starting with the pioneering works  of  \cite{Dowker:1975tf,Hawking:1976ja}; for a more complete list of references see \cite{Elizalde:1994gf}. Much of our exposition and results follows quite closely the vast literature in the subject of functional determinants which has a very solid branch anchored in the more mathematical tradition starting in \cite{Ray:1973sb}; for a more complete list of references see \cite{gilkey1995invariance}. In the bulk of the paper we make an effort to help the interested reader find the original versions of our arguments in the literature. We owe a particularly great debt to  the work of Dunne and Kirsten \cite{Dunne:2006ct} from which we have borrowed even the idea of the title of our manuscript. Our work could be simply described as an extension of theirs to the case of asymptotically $AdS_2$ spacetime rather than flat space. We have, nevertheless, chosen to be as systematic and self-contained as possible in our presentation and results for the benefits of a string-theory oriented exposition.

In order to avoid getting lost in technical details and to highlight our main results for the  benefit of the pragmatic reader, we will present the main results first and postpone their derivation for later.

The paper is organized as follows. In section \ref{sec: results} we summarize the main results of our work, namely, we present $\zeta$-function regularize of radial Laplace-like operators.  In section \ref{sec: examples} we present a number of explicit examples. The systematic derivation of our results is developed in section \ref{sec: derivation}. We conclude in section \ref{sec: conclusions} where we also point out some interesting directions that can be pursued in relations to the current work. 


\section{Main results and discussion}\label{sec: results}
\subsection{Preamble}
Throughout this paper we will work on the disk model of Euclidean $AdS_2$ (or $\mathds{H}_2$) with metric
\begin{empheq}{alignat=7}\label{eq: AdS2}
	ds^2&=L^2\left(d\rho^2+\sinh^2\rho\,d\tau^2\right)\,,
	&\qquad
	\rho&\geq0\,,
	&\quad
	\tau&\sim\tau+2\pi\,.
\end{empheq}
For simplicity we set $L=1$ but we will reinstate the radius in the final expressions. We are interested in Laplace and Dirac-type operators defined in the geometry \eqref{eq: AdS2}  in the presence of additional background fields. Specifically, we consider operators of the form
\begin{empheq}{alignat=7}\label{eq: 2d operator bosons}
	\mathcal{O}&=-g^{\mu\nu}D_{\mu}D_{\nu}+m^2+V\,,
	&\qquad&\textrm{(bosons)}
	\\\label{eq: 2d operator fermions}
	\mathcal{O}&=-i\left(\slashed{D}+\slashed{\partial}\Omega\right)-i\Gamma_{\underline{01}}\left(m+V\right)+W\,,
	&\qquad&\textrm{(fermions)}
\end{empheq}
where the covariant derivative $D_{\mu}=\nabla_{\mu}-iq\mathcal{A}_{\mu}$ includes a $U(1)$ gauge field. Here $m$  and $q$ are arbitrary mass and charge parameters, respectively. It should be clear from the outset that, even though we use the same notation, $m$, $q$, $V$ and  $\mathcal{A}_{\mu}$ need not be the same for bosons and fermions. In the latter case we have included an extra {\it connection}, $d\Omega$   (notice the absence of $i$, thus implying it cannot be gauged away), whose origin is motivated by thinking of these operators as coming from some other geometry that is conformal to $AdS_2$. We also clarify that $W$ and $V$  are not matrix-valued. Rather, they are  scalar functions.

Our goal is to compute the ratio of determinants of the operators \eqref{eq: 2d operator bosons} and \eqref{eq: 2d operator fermions} with the corresponding free operators obtained by setting $\mathcal{A}_{\mu}=\Omega=V=W=0$. For generic choices of the background fields, this is an extremely difficult task and can only be handled on a case by case basis. Considerable progress can be made, however, if one assumes circular symmetry. Consequently, we restrict ourselves to configurations where $\mathcal{A}_{\rho}=0$ and $\mathcal{A}_{\tau}=\mathcal{A}(\rho)$, as well as $V=V(\rho)$, $W=W(\rho)$ and $\Omega=\Omega(\rho)$. The condition $\mathcal{A}_{\rho}=0$ is actually a gauge choice, while the remaining assumptions imply circular symmetry.

A recurring notion in the following sections is the regularity of the eigenfunctions of the operators in question. Accordingly, the background fields must also be regular. Given the topology of $AdS_2$, this translates to
\begin{empheq}{alignat=7}
	\mathcal{A}(\rho)&\underset{\rho\rightarrow0}{\longrightarrow} \rho^{1+\epsilon}\,,
	&\qquad
	\partial_{\rho}\Omega(\rho)&\underset{\rho\rightarrow0}{\longrightarrow}\rho^{\epsilon}\,,
	&\qquad
	\epsilon&\geq0\,,
\end{empheq}
so that the 1-forms $\mathcal{A}(\rho)d\tau$ and $\partial_{\rho}\Omega(\rho)d\rho$ are well-defined at the origin. At infinity the gauge field and connection behave like
\begin{empheq}{alignat=7}
	\mathcal{A}(\rho)&\underset{\rho\rightarrow\infty}{\longrightarrow}\mathcal{A}_{\infty}\,,
	&\qquad
	\partial_{\rho}\Omega(\rho)&\underset{\rho\rightarrow\infty}{\longrightarrow}0\,.
\end{empheq}
On the other hand, the potentials are assumed to decay at least as
\begin{empheq}{alignat=7}
	V(\rho)&\underset{\rho\rightarrow\infty}{\longrightarrow}  \frac{e^{-\rho}}{\rho^{2+\epsilon}}\,
	&\qquad
	W(\rho)&\underset{\rho\rightarrow\infty}{\longrightarrow}\frac{e^{-\frac{\rho}{2}}}{\rho^{1+\epsilon}}\,.
\end{empheq}
Simply put, the background fields must behave in such a way that all the integrals appearing below are finite. These fall-off conditions imply that the operators become effectively free for large $\rho$,
\begin{empheq}{alignat=7}
	\mathcal{O}&\underset{\rho\rightarrow\infty}{\longrightarrow}\mathcal{O}^{\textrm{free}}\,,
\end{empheq}
except for the presence of a constant gauge field, which does not affect in any substantial way the validity of the results.  

The spectral problem at hand is intrinsically two-dimensional but the assumption of circular symmetry reduces it to a one-dimensional calculation. Upon Fourier-transforming the $\tau$ dependence the relevant radial operators become
\begin{empheq}{alignat=7}\label{eq: 1d radial operator bosons}
	\mathcal{O}_l&=-\frac{1}{\sinh\rho}\partial_{\rho}\left(\sinh\rho\,\partial_{\rho}\right)+\frac{\left(l-q\mathcal{A}\right)^2}{\sinh^2\rho}+m^2+V\,,
	&\quad
	l&\in\mathds{Z}\,,
	&\quad&
	\textrm{(bosons)}
	\\\label{eq: 1d radial operator fermions}
	\mathcal{O}_l&=-i\Gamma_{\underline{1}}\left(\partial_{\rho}+\frac{1}{2}\coth\rho+\partial_{\rho}\Omega\right)+\Gamma_{\underline{0}}\frac{\left(l-q\mathcal{A}\right)}{\sinh\rho}-i\Gamma_{\underline{01}}\left(m+V\right)+W\,,
	&\quad
	l&\in\mathds{Z}+\frac{1}{2}\,.
	&\quad&
	\textrm{(fermions)}
\end{empheq}
As a first attempt to reconstruct the full determinant one could write
\begin{empheq}{alignat=7}\label{eq: naive sum}
	\ln\frac{\det\mathcal{O}}{\det\mathcal{O}^{\textrm{free}}}&\overset{?}{=}\sum_{l=-\infty}^{\infty}\ln\frac{\det\mathcal{O}_l}{\det\mathcal{O}_l^{\textrm{free}}}\,.
\end{empheq}
The trouble with this expression, however, is that, even though the ratio $\frac{\det\mathcal{O}_l}{\det\mathcal{O}_l^{\textrm{free}}}$ is well defined, the sum over Fourier modes typically diverges. To give it meaning one could, for example, regulate the sum by imposing a sharp cutoff at $|l|=\Lambda$ and subtract the divergent pieces. In some contexts, an underlying symmetry might even cancel the divergences altogether.  A cutoff regularization, however, might conflict with symmetries of curved spaces, in particular diffeomorphism invariance, rendering this approach not entirely satisfying. A more geometric approach is desirable.

 One would like to insist on the idea of reconstructing the two-dimensional determinants as a product over one-dimensional ones, since the latter are relatively easy to compute. The purpose of this work is to provide a regularization scheme that coincides with the two-dimensional $\zeta$-function formalism, that is,
\begin{empheq}{alignat=7}\label{eq: renormalized determinant}
	\ln\frac{\det\mathcal{O}}{\det\mathcal{O}^{\textrm{free}}}&\equiv-\hat{\zeta}_{\mathcal{O}}'(0)-\ln(\mu^2)\hat{\zeta}_{\mathcal{O}}(0)\,,
	&\qquad
	\hat{\zeta}_{\mathcal{O}}(s)&\equiv\zeta_{\mathcal{O}}(s)-\zeta_{\textrm{free}}(s)\,,
\end{empheq}
where $\mu$ is a mass scale that parametrizes the ambiguity in the renormalization of the determinant. The same definitions apply to the radial operators $\mathcal{O}_l$, although the renormalization scale is absent in one dimension.  For fermions, we define the determinant and $\zeta$-function of the first order operator  in terms of the squared one as
\begin{empheq}{alignat=7}
	\det\mathcal{O}&\equiv\left(\det\mathcal{O}^2\right)^{\frac{1}{2}}\,,
	&\qquad
	\zeta_{\mathcal{O}}(s)&\equiv\frac{1}{2}\zeta_{\mathcal{O}^2}(s)\,.
\end{empheq}
In this context, the correct version of \eqref{eq: naive sum} is
\begin{empheq}{alignat=7}\label{eq: correct sum}
	\zeta_{\mathcal{O}}(s)&=\sum_{l=-\infty}^{\infty}\zeta_{\mathcal{O}_l}(s)\,.
\end{empheq}
This relation is as usual generically not well-defined in the entire complex $s$-plane, only for large enough $\mathrm{Re}\, s$. The problem in the present work then boils down to finding the analytic continuation to $s=0$ of the whole sum and not each individual term separately.


\subsection{Results}
Concerning the bosonic case, our main result is
\begin{empheq}{alignat=7}
	\ln\frac{\det\mathcal{O}}{\det\mathcal{O}^{\textrm{free}}}&=\ln\frac{\det\mathcal{O}_0}{\det\mathcal{O}_0^{\textrm{free}}}+\sum_{l=1}^{\infty}\left(\ln\frac{\det\mathcal{O}_l}{\det\mathcal{O}_l^{\textrm{free}}}+\ln\frac{\det\mathcal{O}_{-l}}{\det\mathcal{O}_{-l}^{\textrm{free}}}+\frac{2}{l}\hat{\zeta}_{\mathcal{O}}(0)\right)-2\left(\ln\left(\mu L\right)+\gamma\right)\hat{\zeta}_{\mathcal{O}}(0)\nn
	\\\label{eq: main result bosons}
	&+\int_0^{\infty}d\rho\,\sinh\rho\ln\left(\frac{\sinh\rho}{2}\right)V-q^2\int_0^{\infty}d\rho\,\frac{\mathcal{A}^2}{\sinh\rho}\,
	\\\nonumber
	\hat{\zeta}_{\mathcal{O}}(0)&=-\frac{1}{2}\int_0^{\infty}d\rho\,\sinh\rho\,V\,,
\end{empheq}
where $\gamma\approx0.57721$ is the Euler-Mascheroni constant. In turn, the ratio of radial determinants for each Fourier mode can be computed as
\begin{empheq}{alignat=7}\label{eq: ratio radial determinants bosons}
	\ln\frac{\det\mathcal{O}_l}{\det\mathcal{O}_l^{\textrm{free}}}&=\lim_{\rho\rightarrow\infty}\ln\frac{\psi_l(\rho)}{\psi_l^{\textrm{free}}(\rho)}\,,
\end{empheq}
where $\psi_l(\rho)$ is the solution to the homogeneous equation for $\mathcal{O}_l$ that is regular at $\rho=0$,
\begin{empheq}{alignat=7}
	\mathcal{O}_l\psi_l&=0\,,
	&\qquad
	\psi_l(\rho)&\underset{\rho\rightarrow0}{\longrightarrow}\rho^{|l|}\,.
\end{empheq}
The normalization   is chosen so that the leading coefficient in the small $\rho$ expansion matches that of the free solution appearing in the denominator\footnote{This is completely analogous to the usual initial conditions $\psi(0)=0$, $\psi'(0)=1$ imposed on the homogeneous functions   appearing in the Gelfand-Yaglom method. In two and higher dimensions, however, the centrifugal barrier implies that the regular solution actually vanishes as a power law depending on the Fourier mode, so $\psi'(0)=1$ must be generalized.} of  \eqref{eq: ratio radial determinants bosons}. 

Similarly, for fermionic operators we get
\begin{empheq}{alignat=7}\nonumber
	\ln\frac{\det\mathcal{O}}{\det{\mathcal{O}^{\textrm{free}}}}&=\sum_{l=\frac{1}{2}}^{\infty}\left(\ln\frac{\det\mathcal{O}_l}{\det\mathcal{O}_l^{\textrm{free}}}+\ln\frac{\det\mathcal{O}_{-l}}{\det\mathcal{O}_{-l}^{\textrm{free}}}+\frac{2}{l+\frac{1}{2}}\hat{\zeta}_{\mathcal{O}}(0)\right)
	-2\left(\ln\left(\mu L\right)+\gamma\right)\hat{\zeta}_{\mathcal{O}}(0)
	\\\nonumber
	&+\int_0^{\infty}d\rho\,\sinh\rho\ln\left(\frac{\sinh\rho}{2}\right)\left(\left(m+V\right)^2-W^2-m^2\right)-q^2\int_0^{\infty}d\rho\,\frac{\mathcal{A}^2}{\sinh\rho} 
	\\
	&-\int_0^{\infty}d\rho\,\sinh\rho\,W^2\,,\label{eq: main result fermions}
	\\\nonumber
	\hat{\zeta}_{\mathcal{O}}(0)&=-\frac{1}{2}\int_0^{\infty}d\rho\,\sinh\rho\left(\left(m+V\right)^2-W^2-m^2\right)\nn\,,
\end{empheq}  
where
\begin{empheq}{alignat=7}\label{eq: ratio radial determinants fermions}
	\ln\frac{\det\mathcal{O}_l}{\det\mathcal{O}_l^{\textrm{free}}}&=\lim_{\rho\rightarrow\infty}\left(\ln\frac{\psi^{(i)}_l(\rho)}{\psi_l^{(i)\,\textrm{free}}(\rho)}+\Omega(\rho)-\Omega(0)\right)\,.
\end{empheq}
Here $\psi^{(i)}_l(\rho)$ is any of the two components of the regular spinor solution to the \emph{first order} homogeneous equation,
\begin{empheq}{alignat=7}\label{homFer}
	\mathcal{O}_l\psi_l&=0\,,
	&\qquad
	\psi_l(\rho)&\underset{\rho\rightarrow0}{\longrightarrow}\rho^{|l|-\frac{1}{2}}\,.
\end{empheq}
The small $\rho$ behavior is displayed only for the leading component\footnote{The other component goes as $\rho^{|l|+\frac{1}{2}}$ with a coefficient that depends on the behavior of the potentials at the origin (see \eqref{lpdnorm}).}. As for  bosons, this component should be normalized so that its   behavior at the origin coincides with that of the free solution to be inserted in \eqref{eq: ratio radial determinants fermions}. We stress that any of the two components can be used in \eqref{eq: ratio radial determinants fermions}.

A few comments are in order. Our results are simple generalizations of those in flat space \cite{Dunne:2006ct}; mainly  replace $\rho\rightarrow\sinh\rho$ for the radial dependence and $\rho\,d\rho\rightarrow\sinh\rho\,d\rho$ in the integration measure. This is related to the fact that, by construction, zeta-function regularization is diffeomorphism invariant, even though expressions \eqref{eq: main result bosons} and \eqref{eq: main result fermions} are written in a particular coordinate system. Also, it is reassuring to check that $\hat{\zeta}_{\mathcal{O}}(0)$ coincides with the general formula in terms of the Seeley coefficient \cite{Drukker:2000ep,Vassilevich:2003xt} (see also appendix \ref{app: Weyl Anomaly})
\begin{empheq}{alignat=7}
	\hat{\zeta}_{\mathcal{O}}(0)&=a_2(1|\mathcal{O})-a_2(1|\mathcal{O}^{\textrm{free}})\,.
\end{empheq}
Another important point is that in an infinite space such as $AdS_2$ there is actually no freedom in choosing the boundary conditions once one imposes that the eigenfunctions are regular everywhere.  An intermediate step in the derivation \eqref{eq: main result bosons} and \eqref{eq: main result fermions} involves putting the system in a finite box of radius $R$ where boundary conditions are indeed relevant. However, the $R\rightarrow\infty$ limit eliminates all traces of these.

As one would expect from circular symmetry, the two-dimensional determinants can be written as a sum of one-dimensional radial determinants. It is important to emphasize, however, that all results are finite and do not require further regularization. It is still useful to compare with the momentum cut-off prescription widely used in context of holographic Wilson loops \cite{Kruczenski:2008zk}\cite{Forini:2015bgo}\cite{Faraggi:2016ekd}\cite{Kim:2012nd}. To that end, we notice that the sums over Fourier modes in \eqref{eq: main result bosons} and \eqref{eq: main result fermions} are rendered finite by the presence of the term $\frac{1}{l}\hat{\zeta}_{\mathcal{O}}(0)$, 
\begin{empheq}{alignat=7}
	\hat{\zeta}_{\mathcal{O}}(0)\sum_{l=1}^{\Lambda}\frac{1}{l}&=\hat{\zeta}(0)\left(\ln\Lambda+\gamma\right)+O(\Lambda^{-1})\,,
\end{empheq}
which cancels a $\ln\Lambda$ divergence in \eqref{eq: naive sum}. It was not obvious a priori that the correct coefficient was $\hat{\zeta}_{\mathcal{O}}(0)$. In the fermionic case, it is also crucial to include the $\Omega$ term in \eqref{eq: ratio radial determinants fermions} so that the sum is free of linear divergences. In retrospect, this justifies the rescaling of the boundary conditions done in \cite{Faraggi:2016ekd}. Finally, zeta-function regularization systematically fixes all the finite terms in \eqref{eq: main result bosons} and \eqref{eq: main result fermions} that depend explicitly on the background fields, which a cut-off method could not possibly foresee.


\subsection{Conformal $AdS_2$ spaces}
\label{subsec: conformal AdS}

A simple generalization of the methods presented here include functional determinants defined on spaces that are conformally equivalent to $AdS_2$, namely,
\begin{empheq}{alignat=7}
	\label{conf}
	ds_M^2&=Mds^2\,,
\end{empheq}
where the conformal factor $M$ is smooth everywhere so as to not change the topology\footnote{Of course, any two-dimensional geometry is conformally equivalent to any other two-dimensional geometry. This is, however, a local statement. The emphasis here is that the conformal factor does not blow up anywhere so the topology is still that of a disk.}. The Laplace and Dirac operators in the two geometries, are related by using  
\begin{empheq}{alignat=7}
	e^{\underline{a}}_{_M}=\sqrt{M}e^{\underline{a}},\qquad w^{\underline{ab}}_{_ M} &=w^{\underline{ab}}    -\frac{1}{2M}\left(\partial^{\underline{a}}Me^{\underline{b}}-\partial^{\underline{b}}Me^{\underline{a}}\right)\,,
\end{empheq}
where $\partial^{\underline{a}}M=e^{\underline{a}\mu}\partial_\mu M$ and $e^{\underline{a}}_\mu e_{\underline{b}}^\mu=\delta^{\underline{a}}_{\underline{b}}$. Some Dirac matrix algebra then shows 
\begin{empheq}{alignat=7}\label{spinc}
	\nabla^2_{_M}=\frac1M \nabla^2,\qquad \slashed{\nabla}\!_{ _M}&=\frac{1}{\sqrt{M}}\left(\slashed{\nabla}  +\frac{\slashed{\partial}M}{4M}\right)\,.
\end{empheq}
This leads us to consider more general operators of the form
\begin{empheq}{alignat=7}\label{eq: rescaled operators bosons}
	\mathcal{O}_M&=M^{-1}\mathcal{O}\,,
	&\qquad
	\mathcal{O}&=-g^{\mu\nu}D_{\mu}D_{\nu}+m^2+V\,,
	&\qquad&
	\textrm{(bosons)}
	\\\label{eq: rescaled operators fermions}
	\mathcal{O}_M&=M^{-\frac{1}{2}}\mathcal{O}\,,
	&\qquad
	\mathcal{O}&=-i\left(\slashed{D}+\slashed{\partial}\Omega\right)-i\Gamma_{\underline{01}}\left(m+V\right)+W\,,
	&\qquad&
	\textrm{(fermions)}
\end{empheq}
where $\mathcal{O}$ is defined in the $AdS_2$ geometry as before. Notice that any potential terms originally appearing in $\mathcal{O}_M=-D^2_M+\cdots$ or $\mathcal{O}_M=-i\slashed{D}_M+\cdots$ will need to be rescaled by $M$ or $M^{\frac{1}{2}}$ in order to write them in this fashion.  In the fermionic case there is an additional contribution $\frac{1}{4}\slashed{\partial}\ln M$ coming from the spin connection in \eqref{spinc}, which we have absorbed in $\slashed{\partial}\Omega$. As before we assume that the conformal factor depends only on the radial coordinate; circular symmetry would otherwise be lost. The gauge field is unaffected by the rescaling. 

The determinants of $\mathcal{O}_M$ and $\mathcal{O}$ are connected by the standard Weyl anomaly calculation (see appendix \ref{app: Weyl Anomaly}). Taking the ratio with the free operator on $AdS_2$ we find
\begin{empheq}{alignat=7}\label{eq: Weyl anomaly ratio bosons}
	\ln\left(\frac{\det\mathcal{O}_{M}}{\det\mathcal{O}^{\textrm{free}}}\right)&=\ln\left(\frac{\det\mathcal{O}}{\det\mathcal{O}^{\textrm{free}}}\right)+\frac{1}{4\pi}\int d^2\sigma\sqrt{g}\,\ln M\left[m^2+V-\frac{1}{6}R+\frac{1}{12}\nabla^2\ln M\right]
\end{empheq}
for bosons, while for fermions the anomaly reads
\begin{empheq}{alignat=7}\label{eq: Weyl anomaly ratio fermions}
	\ln\left(\frac{\det\mathcal{O}_{M}}{\det\mathcal{O}^{\textrm{free}}}\right)&=\ln\left(\frac{\det\mathcal{O}}{\det\mathcal{O}^{\textrm{free}}}\right)+\frac{1}{4\pi}\int d^2\sigma\sqrt{g}\,\ln M\left[\left(m+V\right)^2-W^2+\frac{1}{12}R-\frac{1}{24}\nabla^2\ln M\right]\,.
\end{empheq}
In each expression the first term on the right hand side can be computed using the results of the previous section. The second term accounts for the rescaling. We have assumed that $M\rightarrow1$ as $\rho\rightarrow\infty$ so the space is asymptotically $AdS_2$, which explains the absence of boundary terms.


\section{Examples}\label{sec: examples}
In this section we apply the methods developed here to two examples borrowed from the literature on holographic Wilson loops \cite{Drukker:2007qr,Forini:2015bgo,Faraggi:2016ekd}. See also \cite{Aguilera-Damia:2018twq}.
\subsection{Bosons}
For the bosonic case we take
\begin{empheq}{alignat=7}
	\mathcal{O}_{M}&=M^{-1}\mathcal{O}\,,
	&\qquad
	\mathcal{O}&=-g^{\mu\nu}D_{\mu}D_{\nu}+V\,,
	\qquad
	D_{\mu}=\nabla_{\mu}+i\mathcal{A}_{\mu}\,,
\end{empheq}
with
\begin{empheq}{alignat=7}
	M(\rho)&=1+\frac{\sin^2\theta(\rho)}{\sinh^2\rho}\,,
	&\qquad
	\mathcal{A}(\rho)&=1-\frac{1+\cosh\rho\cos\theta(\rho)}{\cosh\rho+\cos\theta(\rho)}\,,
	&\qquad
	V(\rho)=-\frac{\partial_{\rho}\mathcal{A}(\rho)}{\sinh\rho}\,.
\end{empheq}
The function $\theta(\rho)$ is given by
\begin{empheq}{alignat=7}
	\sin\theta(\rho)&=\frac{\sinh\rho\sin\theta_0}{\cosh\rho+\cos\theta_0}\,,
\end{empheq}
where $0\leq\theta_0\leq\frac{\pi}{2}$ is a parameter. The free operator corresponds to
\begin{empheq}{alignat=7}
	\mathcal{O}^{\textrm{free}}&\equiv \mathcal{O}|_{_{\theta_0=0}}=\mathcal{O}_{M}|_{_{\theta_0=0}}=-\nabla^2\,.
\end{empheq}

Let us use our result \eqref{eq: main result bosons} to compute the ratio of determinants between $\mathcal{O}$ and $\mathcal{O}^{\textrm{free}}$. We will include the effect of the Weyl anomaly in \eqref{eq: Weyl anomaly ratio bosons} at the end. First,
\begin{empheq}{alignat=7}
	\hat{\zeta}_{\mathcal{O}}(0)&=-\frac{1}{2}\int_0^{\infty}d\rho\,\sinh\rho\,V
	\\\nonumber
	&=\sin^2\frac{\theta_0}{2}\,.
\end{empheq}
Similarly,
\begin{empheq}{alignat=7}
	\int_0^{\infty}d\rho\,\sinh\rho\,\ln\left(\frac{\sinh\rho}{2}\right)V&=-\frac{1}{2}\theta_0\sin\theta_0+\cos\theta_0\ln\cos\frac{\theta_0}{2}\,,
\end{empheq}
and
\begin{empheq}{alignat=7}
	\int_0^{\infty}d\rho\,\frac{\mathcal{A}^2}{\sinh\rho}&=-\sin^2\frac{\theta_0}{2}-2\ln\cos\frac{\theta_0}{2}\,.
\end{empheq}
Next, notice that the general solution to the differential equation
\begin{empheq}{alignat=7}\label{bosCount}
	\mathcal{O}_l\psi_l&=0\,,
	&\qquad
	\mathcal{O}_l&=-\frac{1}{\sinh\rho}\partial_{\rho}\left(\sinh\rho\,\partial_{\rho}\right)+\frac{\left(l+\mathcal{A}\right)^2}{\sinh^2\rho}-\frac{\partial_{\rho}\mathcal{A}}{\sinh\rho}\,,
	&\qquad
	l&\in\mathds{Z}\,,
\end{empheq}
is
\begin{empheq}{alignat=7}
	\psi_l(\rho)&=\left(\tanh\frac{\rho}{2}\right)^{-l}e^{-\mathcal{W}(\rho)}\left(C_1+C_2\int d\rho\,\left(\tanh\frac{\rho}{2}\right)^{2l}\frac{e^{2\mathcal{W}(\rho)}}{\sinh\rho}\right)\,,
	&\qquad
	\partial_{\rho}\mathcal{W}(\rho)&=\frac{\mathcal{A}(\rho)}{\sinh\rho}\,.
\end{empheq}
Since $\mathcal{W}(\rho)$ is finite at $\rho=0$, we see that for $l<0$ the regular solution corresponds to $C_2=0$, whereas for $l>0$ we must set $C_1=0$. Making sure that the normalization is the same as for the free solution we find
\begin{empheq}{alignat=7}
	\psi_l(\rho)&=\left\{
	\begin{array}{cc}
		{\displaystyle\frac{\cos\frac{\theta_0}{2}\left(2\tanh\frac{\rho}{2}\right)^{-l}\left(\cosh\rho+1\right)}{\sqrt{\cosh^2\rho+2\cosh\rho\cos\theta_0+1}}} & l\leq0
		\\\\
		{\displaystyle\frac{\left(2\tanh\frac{\rho}{2}\right)^l\sqrt{\cosh^2\rho+2\cosh\rho\cos\theta_0+1}}{(l+2)\cos\frac{\theta_0}{2}\left(\cosh\rho+1\right)}\left(l+\frac{2\left(\cosh\rho+1\right)^2\cos^2\frac{\theta_0}{2}}{\cosh^2\rho+2\cosh\rho\cos\theta_0+1}\right)} & l\geq0
	\end{array}
	\right.\,.
\end{empheq}
Thus,
\begin{empheq}{alignat=7}
	\ln\frac{\det\mathcal{O}_l}{\det\mathcal{O}_l^{\textrm{free}}}&=\left\{
	\begin{array}{cc}
		{\displaystyle\ln\cos\frac{\theta_0}{2}} & l\leq0
		\\\\
		{\displaystyle-\ln\cos\frac{\theta_0}{2}+\ln\left(\frac{l+2\cos^2\frac{\theta_0}{2}}{l+2}\right)} & l\geq0
	\end{array}
	\right.\,.
\end{empheq}
Happily, the sum over Fourier modes can be computed in closed form. Indeed,
\begin{empheq}{alignat=7}
	\sum_{l=1}^{\infty}\left(\ln\frac{\det\mathcal{O}_l}{\det\mathcal{O}_l^{\textrm{free}}}+\ln\frac{\det\mathcal{O}_{-l}}{\det\mathcal{O}_{-l}^{\textrm{free}}}+\frac{2}{l}\hat{\zeta}_{\mathcal{O}}(0)\right)&=\sum_{l=1}^{\infty}\left(\ln\left(\frac{l+2\cos^2\frac{\theta_0}{2}}{l+2}\right)+\frac{2}{l}\sin^2\frac{\theta_0}{2}\right)\nn
	\\
	&=-\ln\Gamma\left(2\cos^2\frac{\theta_0}{2}\right)-2\ln\cos\frac{\theta_0}{2}+2\gamma\sin^2\frac{\theta_0}{2}\,.
\end{empheq}
Notice that were it not for the $\hat{\zeta}_{\mathcal{O}}(0)$-term, the sum would have been divergent, which is precisely the situation faced in \cite{Kruczenski:2008zk,Forini:2015bgo,Faraggi:2016ekd}. Putting everything together we arrive at
\begin{empheq}{alignat=7}\label{eq: Wilson bosons AdS}
	\ln\frac{\det\mathcal{O}}{\det\mathcal{O}^{\rm free}}&=-\ln\Gamma\left(2\cos^2\frac{\theta_0}{2}\right)+2\cos^2\frac{\theta_0}{2}\ln\cos\frac{\theta_0}{2}+\sin^2\frac{\theta_0}{2}-\frac{1}{2}\theta_0\sin\theta_0\nn
	\\
	&=-\frac{\gamma}{2}\theta_0^2+\left(\frac{19}{96}+\frac{\gamma}{24}-\frac{\pi^2}{48}\right)\theta_0^4+O\left(\theta_0^6\right)\,,
\end{empheq}
where we have set $\mu=1$ for simplicity. Finally, we compute the Weyl anomaly relating the determinants of $\mathcal{O}_M $ and $\mathcal{O} $. It reads
\begin{empheq}{alignat=7}
	\frac{1}{4\pi}\int d^2\sigma\sqrt{g}\,\ln M\left[V-\frac{1}{6}R+\frac{1}{12}\nabla^2\ln M\right]&=\left(\frac{1}{3}+2\cos^2\frac{\theta_0}{2}\right)\ln\cos\frac{\theta_0}{2}-\frac{1}{2}\sin^2\frac{\theta_0}{2}+\frac{1}{2}\theta_0\sin\theta_0\,.
\end{empheq}
Combining this with the previous expression we find
\begin{empheq}{alignat=7}\label{eq: Wilson bosons}
	\ln\frac{\det\mathcal{O}_M}{\det\mathcal{O}^{\rm free}}&=-\ln\Gamma\left(2\cos^2\frac{\theta_0}{2}\right)+\left(\frac{1}{3}+4\cos^2\frac{\theta_0}{2}\right)\ln\cos\frac{\theta_0}{2}+\frac{1}{2}\sin^2\frac{\theta_0}{2}\nn
	\\
	&=\left(\frac{1}{12}-\frac{\gamma}{2}\right)\theta_0^2+\left(\frac{101}{576}+\frac{\gamma}{24}-\frac{\pi^2}{48}\right)\theta_0^4+O\left(\theta_0^6\right)\,.
\end{empheq}

The reason we have expanded our results for small $\theta_0$ is to compare them against the perturbative technique developed in \cite{Forini:2017whz}. While we spare the details of the calculation, we confirm that the leading terms in \eqref{eq: Wilson bosons AdS} and \eqref{eq: Wilson bosons} are in fact reproduced, independently, by this method. It would be interesting to extend the perturbative method to next order in the expansion parameter and check that it also reproduces the $O\left(\theta_0^4\right)$ terms.
\subsection{Fermions} \label{fermexample}
As a fermionic example we consider the operator
\begin{empheq}{alignat=7}
	\mathcal{O}_M&=M^{-\frac{1}{2}}\mathcal{O}\,,
	&\qquad
	\mathcal{O}&=-i\left(\slashed{D}+\frac{1}{4}\slashed{\partial}\ln M\right)-i\Gamma_{\underline{01}}\left(1+V\right)+W\,,
	\qquad
	D_{\mu}=\nabla_{\mu}+\frac{i}{2}\mathcal{A}_{\mu}\,,
\end{empheq}
where $M(\rho)$ and $\mathcal{A}(\rho)$ are the same as before and
\begin{empheq}{alignat=7}
	V(\rho)=\frac{1}{\sqrt{M(\rho)}}-1\,,
	&\qquad
	W(\rho)&=\frac{\sin^2\theta(\rho)}{\sqrt{M(\rho)}\sinh^2\rho}\,.
\end{empheq}
The free operator reads
\begin{empheq}{alignat=7}
	\mathcal{O}^{\textrm{free}}&=\mathcal{O} |_{_{\theta_0=0}}=\mathcal{O}_M |_{_{\theta_0=0}}=-\slashed{\nabla}-i\Gamma_{\underline{01}}\,.
\end{empheq}

This time the relevant formulas are \eqref{eq: main result fermions} and \eqref{eq: Weyl anomaly ratio fermions}. We find
\begin{empheq}{alignat=7}
	\hat{\zeta}_{\mathcal{O}}(0)&=-\frac{1}{2}\int_0^{\infty}d\rho\,\sinh\rho\left(\left(m+V\right)^2-m^2-W^2\right)
	\\
	&=\sin^2\frac{\theta_0}{2}\,,
\end{empheq}
as well as
\begin{empheq}{alignat=7}
	\int_0^{\infty}d\rho\,\sinh\rho\ln\left(\frac{\sinh\rho}{2}\right)\left(\left(m+V\right)^2-W^2-m^2\right)&=2\cos\theta_0\ln\cos\frac{\theta_0}{2}\,,
\end{empheq}
together with
\begin{empheq}{alignat=7}
	\int_0^{\infty}d\rho\,\sinh\rho\,W^2&=2\sin^2\frac{\theta_0}{2}-\frac{1}{2}\theta_0\sin\theta_0\,,
\end{empheq}
and
\be
  \lim_{\rho\to\infty}\left(\Omega(\rho)-\Omega(0)\right)=\lim_{\rho\to\infty}\frac14\ln\left(\frac{M(\rho)}{M(0)}\right)=\frac12\ln\cos\frac{\theta_0}{2}
\ee
The integral involving the gauge field is the same as in the bosonic example. Solving the differential equation, however, is more involved in this case given the spinor structure of the fields. The radial problem is
\begin{empheq}{alignat=7}
	\mathcal{O}_l\psi_l&=0\,,
	&\qquad
	\mathcal{O}_l&=-i\sigma_1\left(\partial_{\rho}+\frac{1}{2}\coth\rho+\frac{1}{4}\partial_{\rho}\ln M\right)-\frac{1}{\sinh\rho}\sigma_2\left(l+\frac{1}{2}\mathcal{A}\right)+\sigma_3\left(1+V\right)+W\,,
\end{empheq}
with $l\in\mathds{Z}+\frac{1}{2}$. Letting
\begin{empheq}{alignat=7}
	\psi_l(\rho)&=\left(
	\begin{array}{c}
		u_l(\rho)
		\\
		v_l(\rho)
	\end{array}
	\right)\,,
\end{empheq}
we can solve algebraically for $u_l(\rho)$ to find\footnote{Notice that $M=\left(1+V+W\right)^2$ which  considerably simplifies the calculations.}
\begin{empheq}{alignat=7}\label{ferCount}
	-\frac{1}{\sinh\rho}\partial_{\rho}\left(\sinh\rho\,\partial_{\rho}v_l(\rho)\right)+\frac{\left(l+\mathcal{B}\right)^2}{\sinh^2\rho}v_l(\rho)-\frac{\partial_{\rho}\mathcal{B}}{\sinh\rho}v_l(\rho)&=0\,,
\end{empheq}
where
\begin{empheq}{alignat=7}
	\mathcal{B}&=\frac{1}{2}\mathcal{A}-\sinh\rho\left(\frac{1}{2}\coth\rho+\frac{1}{4}\partial_{\rho}\ln M\right)\,.
\end{empheq}
Equation \eqref{ferCount} has the same form as its bosonic counterpart \eqref{bosCount}, but we write its general solution slightly differently,
\begin{empheq}{alignat=7}
	v_l(\rho)&=\left(\tanh\frac{\rho}{2}\right)^{-l+\frac{1}{2}}e^{-\mathcal{W}(\rho)}\left(C_1+C_2\int d\rho\,\left(\tanh\frac{\rho}{2}\right)^{2l-1}\frac{e^{2\mathcal{W}(\rho)}}{\sinh\rho}\right)\,,
	&\qquad
	\partial_{\rho}\mathcal{W}(\rho)&=\frac{\mathcal{B}(\rho)+\frac{1}{2}}{\sinh\rho}\,.
\end{empheq}
When defined in this way, the prepotential $\mathcal{W}$ is finite at $\rho=0$, making the analysis simpler. We then get
\begin{empheq}{alignat=7}
	u^{(-)}_l(\rho)&=\frac{\left(2\tanh\frac{\rho}{2}\right)^{-l-\frac{1}{2}}}{\left(l-\frac{1}{2}\right)}\sqrt{\frac{2\left(\cosh\rho+\cos\theta_0\right)}{\cosh^2\rho+2\cosh\rho\cos\theta_0+1}}\left(l+\frac{1}{2}-\frac{\cosh^2\rho+2\cosh\rho\cos\theta_0+1}{2\left(\cosh\rho+\cos\theta_0\right)}\right)\,,
	\\
	v^{(-)}_l(\rho)&=\frac{i\left(2\tanh\frac{\rho}{2}\right)^{-l-\frac{1}{2}}\sinh\rho}{2\left(l-\frac{1}{2}\right)}\sqrt{\frac{2}{\cosh\rho+\cos\theta_0}}\,,
\end{empheq}
for $l\leq-\frac{1}{2}$, and
\begin{empheq}{alignat=7}
	u^{(+)}_l(\rho)=&\frac{i\left(2\tanh\frac{\rho}{2}\right)^{l+\frac12}}{2\cos\frac{\theta_0}{2}\sqrt{\left(\cos\theta_0+\cosh\rho\right)\left(1+2\cos\theta_0\cosh\rho+\cosh^2\rho\right)}(2l+1)(2l+3)} \nn\\
	 &\qquad \quad \times \left(2\cos\theta_0+(2l+1)\cos\theta_0+\cosh\rho\left(2l+1+2\cos\theta_0\right)\left(2\cos\theta_0+\cosh\rho\right)\right)\,,
\end{empheq}
\begin{empheq}{alignat=7}
	v^{(+)}_l(\rho)=&\frac{\left(2\tanh\frac{\rho}{2}\right)^{l+\frac12}(2l-1)}{\cos\frac{\theta_0}{2}\sinh\rho\sqrt{2\left(\cos\theta_0+\cosh\rho\right)}(2l+1)}\nn\\
	& \qquad \qquad \qquad \qquad \times\left(\cos\theta_0+\frac{\left(2l+1+\cos\theta_0\right)\left(1+(2l+1)\cosh\rho+\cosh^2\rho\right)}{(2l-1)(2l+3)}\right)\,,
\end{empheq}
for $l\geq\frac{1}{2}$. The overall normalization constants have been chosen so that the behavior at the origin coincides with \eqref{lpdnorm} for $l\geq\frac{1}{2}$ and \eqref{lpdnorm2} for $l\leq-\frac{1}{2}$. 

Expanding for $\rho\to\infty$ and making the quotient with the free solutions we can compute the sum over Fourier modes, which yields
\begin{align} 
\sum_{l=\frac{1}{2}}^{\infty}\left(\ln\frac{\det\mathcal{O}_l}{\det\mathcal{O}_l^{\textrm{free}}}+\ln\frac{\det\mathcal{O}_{-l}}{\det\mathcal{O}_{-l}^{\textrm{free}}}+\frac{2}{l+\frac{1}{2}}\hat{\zeta}_{\mathcal{O}}(0)\right)&=\sum_{l=\frac12}^{\infty}\left(\ln\left(\frac{l+\frac12+\cos\theta_0}{l+\frac32}\right)+\frac{2}{l+\frac12}\sin^2\frac{\theta_0}{2}\right)\nn\\
&=-\ln\Gamma\left(2\cos^2\frac{\theta_0}{2}\right)+2 \gamma\,\sin^2\frac{\theta_0}{2} \,.
\end{align} 
Note that the sum is rendered finite due to the presence of both the $\hat{\zeta}_{\mathcal{O}}(0)$ and the $\left(\Omega(\infty)-\Omega(0)\right)$ terms. 

Collecting all the pieces, we finally obtain
\begin{align}
\ln\frac{\det\mathcal{O}}{\det\mathcal{O}^{\rm free}}&=-\ln \Gamma\left(2\cos^2\frac{\theta_0}{2}\right)+\left(\frac12+2\cos\theta_0\right)\ln\cos\frac{\theta_0}{2}-\frac74\sin^2\frac{\theta_0}{2}+\frac{\theta_0}{2}\sin\theta_0\nn  \\ 
&= \frac12\left(\frac12-\gamma\right)\theta_0^2 + \frac{1}{384}\left(57+16 \gamma -8\pi^2\right)\theta_0^4 \, + \, O\left(\theta_0^6\right)\,,
\end{align}
where we have set $\mu=1$ for simplicity. In order to obtain the determinant of $\mathcal{O}_M(\theta_0)$, we still have to compute the Weyl anomaly contribution, which in this case reads
\be
\frac{1}{4\pi}\int d^2\sigma\sqrt{g}\,\ln M\left[\left(m+V\right)^2-W^2+\frac{1}{12}R-\frac{1}{24}\nabla^2\ln M\right]= \frac74\sin^2\frac{\theta_0}{2}+\frac{11}{6}\ln\cos\frac{\theta_0}{2}
\ee
thus arriving to the following expression 
\begin{align}
\ln\frac{\det\mathcal{O}_M}{\det\mathcal{O}^{\rm free}}&=-\ln \Gamma\left(2\cos^2\frac{\theta_0}{2}\right)+2\cos\theta_0\ln\cos\frac{\theta_0}{2}+\frac73\ln\cos\frac{\theta_0}{2}
+\frac{\theta_0}{2}\sin\theta_0 \nn\\
&= \frac12\left(\frac{11}{12}-\gamma\right)\theta_0^2 \, + \, \frac{1}{576}\left(59+24\gamma-12\pi^2\right)\theta_0^4\, + \, O\left(\theta_0^6\right)\,.\label{perturbative fermions}
\end{align}
Note the first term is in perfect agreement with the perturbative result reported in \cite{Forini:2017whz}. As in the bosonic case, it would be interesting to check the next order in \eqref{perturbative fermions} by extending the perturbative analysis proposed in \cite{Forini:2017whz} up to $O\left(\theta_0^4\right)$.


\section{Derivation}\label{sec: derivation}
Having discussed the results of the paper and some simple examples, in this section we provide a detailed derivation of equations \eqref{eq: main result bosons} and \eqref{eq: main result fermions}. The procedure essentially mimics the approach taken for flat space in \cite{Dunne:2006ct}.  For the treatment of fermionic determinants we follow \cite{Bordag:1996fv,Bordag:1998tg}. We point the reader to these references for any omitted details, although we do try to make the discussion self-contained. See also \cite{Kirsten:2000ad,Dunne:2006ct,Bordag:1998tg,Elizalde:1994gf,Kirsten:2000xc}.

The main goal is to find  the analytic continuation of expression \eqref{eq: correct sum} to $s=0$.   This is achieved in three steps: i) finding a useful integral representation of the  radial zeta functions using scattering data; ii) give meaning   to the sum over   Fourier modes when evaluated at $s=0$ by an appropriate subtraction; iii)   analytically continue the subtracted terms via Riemann zeta-function.

Before we proceed, a brief comment on notation. It is customary to parametrize the eigenvalues of the $AdS_2$ operators \eqref{eq: 2d operator bosons} and \eqref{eq: 2d operator fermions} by
\begin{empheq}{alignat=7}\label{eq: eigenvalues radial momentum}
	\lambda(\nu)&=\nu^2+\nu_0^2\,,
	&\qquad
	\nu_0&=\sqrt{\frac{1}{4}+m^2}\,,
	&\qquad\textrm{(bosons)}
	\\
	\lambda(\nu)&=\pm\sqrt{\nu^2+\nu_0^2}\,,
	&\qquad
	\nu_0&=m\,,
	&\qquad\textrm{(fermions)}
\end{empheq}
and we adhere to this notation through the rest of the paper. As will become clear below, the variable $\nu$ has the interpretation of a radial momentum.



\subsection{$\zeta$-function as a contour integral}

Consider the bosonic operator \eqref{eq: 2d operator bosons}. We assume it to be Hermitian and  positive definite. Suppose for the moment that the eigenvalues are discrete. This can be achieved by putting the system in a finite spherical box of radius $R$ and eventually taking $R\rightarrow\infty$. For simplicity, we exclude the possibility of zero modes. The spectrum then consists of a finite number of (bound) states with $0<\lambda<\nu_0^2$ and an infinite number of (scattering) states with $\lambda>\nu_0^2$. The zeta-function is symbolically defined as
\begin{empheq}{alignat=7}\label{eq: definition zeta-function}
	\zeta_{\mathcal{O}}(s)&\equiv\sum_n\lambda_n^{-s}\,,
\end{empheq}
where $n$ runs over the full spectrum. Although obviously not valid at $s=0$, this representation of $\zeta_{\mathcal{O}}(s)$ does have meaning in regions of the complex $s$-plane where the sum converges\footnote{If $\lambda_n\sim n^k$, $k>0$ for $n\rightarrow\infty$, then $\textrm{Re}\,s>\frac{1}{k}$.}, and motivates the definition \eqref{eq: renormalized determinant} of the regularized determinant\footnote{The mass scale $\mu$ appears because of the rescaling $\lambda\rightarrow\mu^2\lambda$ needed to make the eigenvalues dimensionless in \eqref{eq: definition zeta-function}.}. However, in order to compute the quantities $\zeta_{\mathcal{O}}(0)$ and $\zeta'_{\mathcal{O}}(0)$ one must first analitically continue the sum to an expression that is well-defined at the origin. Precisely, the main objective in this  section is to provide the details of the continuation procedure for operators in $AdS_2$ displaying circular symmetry. Under these conditions the spectral problem is separable and the zeta-function can always be written as
\begin{empheq}{alignat=7}\label{eq: zeta-function circular symmetry}
	\zeta_{\mathcal{O}}(s)&=\sum_{l\,\in\,\mathds{Z}}\zeta_{\mathcal{O}_l}(s)\,,
	&\qquad
	\zeta_{\mathcal{O}_l}(s)&\equiv\sum_i\lambda_{(l,i)}^{-s}\,,
\end{empheq}
where $i$ labels the eigenvalues of the radial operators $\mathcal{O}_l$ given in \eqref{eq: 1d radial operator bosons}. In general, it is not enough to simply continue $\zeta_{\mathcal{O}_l}(s)$ to $s=0$ and then sum over Fourier modes since the resulting series will be divergent.

\begin{figure}[h]
\begin{center}
\includegraphics[width=0.45\textwidth]{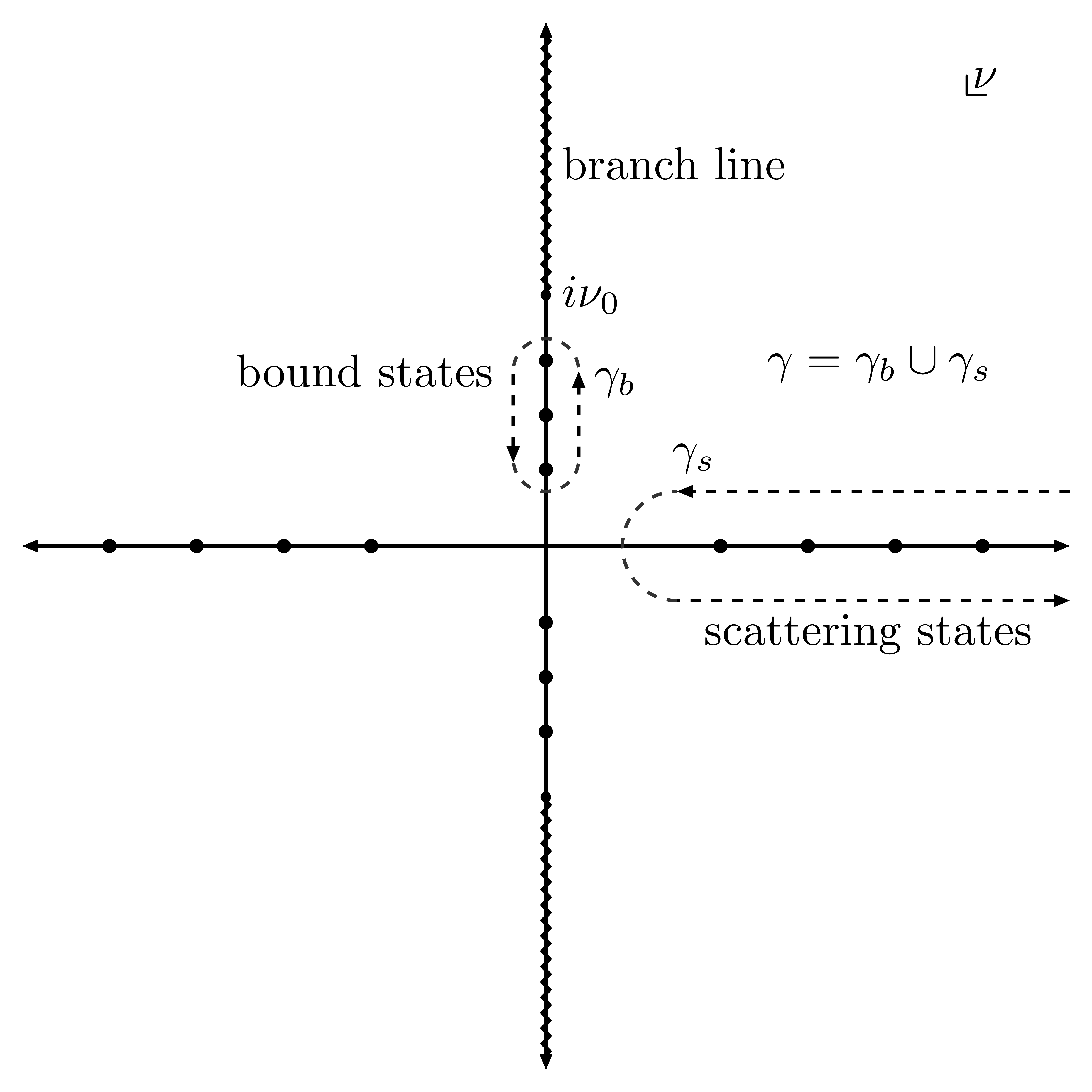}
\includegraphics[width=0.45\textwidth]{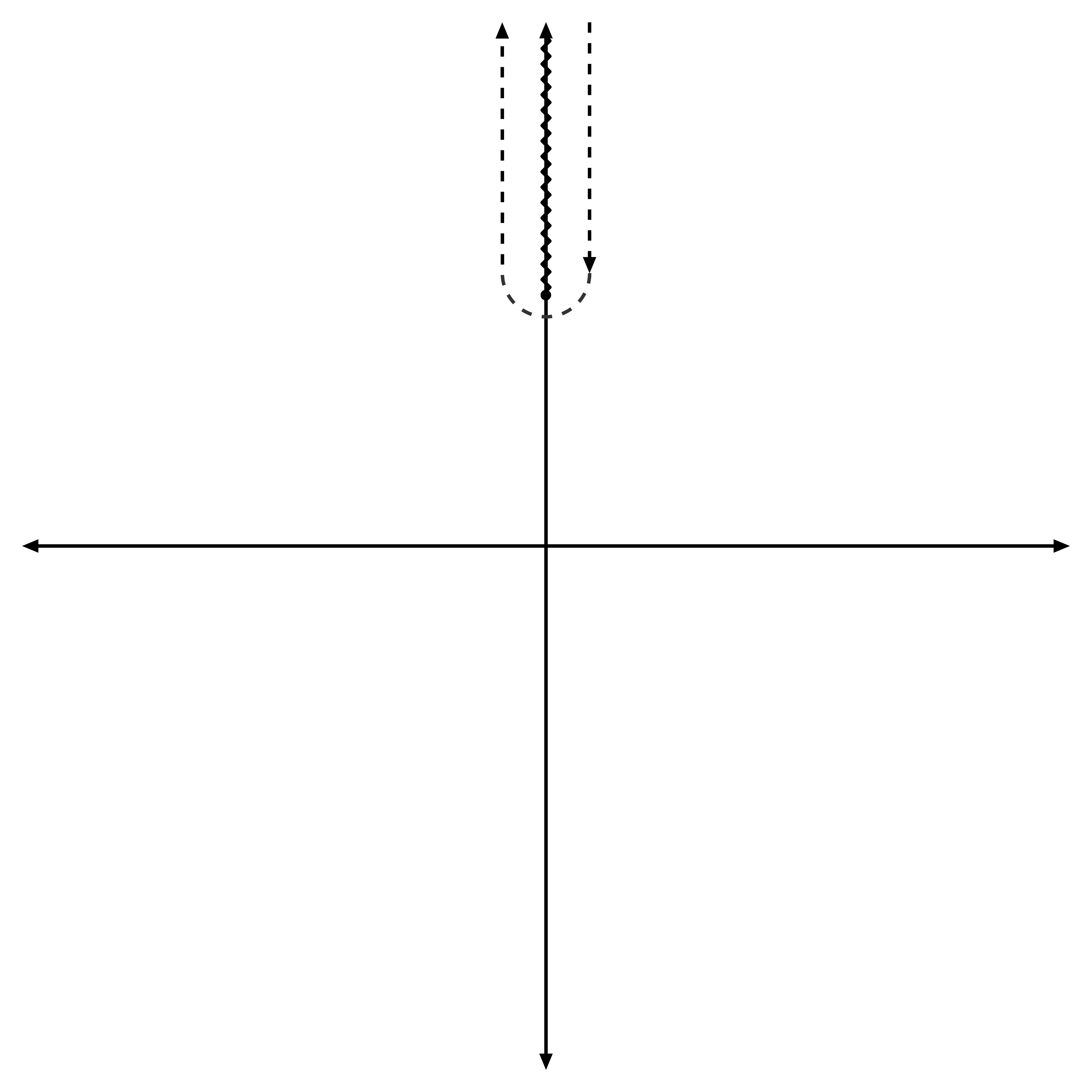}
\end{center}
\label{fig: contour}
\caption{Left: contour in the complex $\nu$-plane for the integral \eqref{zetaC}. Right: after deforming the contour, the integral is performed over the branch cut at the positive imaginary axis.}
\end{figure}

The first step is to find a more suitable representation of the zeta-function. This can be done by trading the sum over $i$ in \eqref{eq: zeta-function circular symmetry} for a contour integral via the residue theorem. In terms of the momentum $\nu$ introduced in \eqref{eq: eigenvalues radial momentum}, the zeta-function for the radial operators can be written as \cite{Kirsten:2007ev}
\begin{empheq}{alignat=7}\label{zetaC}
	\zeta_{\mathcal{O}_l}(s)&=\oint_{\gamma}\frac{d\nu}{2\pi i}\left(\nu^2+\nu_0^2\right)^{-s}\partial_{\nu}\ln f_l(\nu)\,,
\end{empheq}
where $f_l(\nu)$ is a holomorphic function that has simple zeros at the location of the eigenvalues $\lambda_{(l,i)}=\nu^2_{(l,i)}+\nu_0^2$ and $\gamma$ is a path enclosing them all (see figure \ref{fig: contour}).
The logarithm is there to ensure that the residue at each pole is equal to $1$. How do we find such a function $f_l(\nu)$? Imagine solving the differential equation $\mathcal{O}_l\psi=\lambda(\nu)\psi$. Being second order, it will have two independent solutions. These will depend on $\nu$, which at this point is an unspecified parameter. The first consideration we need to make is that we restrict the spectral problem to functions that are smooth everywhere. In particular, for  $AdS_2$, this means regularity\footnote{Moreover, near the origin the operator reduces to that in flat space and the $AdS$ features become irrelevant.} at $\rho=0$. Up to an overall normalization, there is a unique solution satisfying this requirement. Call it $\phi_{(l,\nu)}(\rho)$. The second observation is that the actual eigenvalues are determined by the boundary conditions. For the Dirichlet case, for example, we impose
$\phi_{(l,\nu)}(R)=0$.  This relation should be understood as an equation for $\nu$, having in general infinitely many solutions $\nu=\nu_{(l,i)}$. Extending the domain to the entire complex $\nu$-plane, we identify $f_l(\nu)\equiv\phi_{(l,\nu)}(R)$. Indeed, this function has a simple zero whenever $\nu$ corresponds to one of the eigenvalues of the operator $\mathcal{O}_l$.

The countour integral can be manipulated using  standard techniques of complex analysis. To that end, notice that the function $\left(\nu^2+\nu_0^2\right)^{-s}$ has branch points at $\nu=\pm i\nu_0$. We choose to place the branch cuts along the imaginary axis, as shown in figure \ref{fig: contour}. Taking into account the symmetry $\nu\rightarrow-\nu$ we can deform the path so that it surrounds one of the cuts. The integrand then picks up a phase $e^{\pm i\pi s}$ on each side of the cut and we find
\begin{empheq}{alignat=7}
	\zeta_{\mathcal{O}_l}(s)&=\frac{\sin\pi s}{\pi}\int_{\nu_0}^{\infty}d\nu\,\left(\nu^2-\nu_0
^2\right)^{-s}\partial_{\nu}\ln\phi_{(l,i\nu)}(R)\,.
\end{empheq}
The above representation of the zeta-function is typically not defined at $s=0$ due to the large $\nu$ behavior of $\phi_{(l,i\nu)}$, and its analytic continuation will depend on the details of the operator at hand. 

The behavior improves if we subtract the contribution of some reference (free/solvable) operator\footnote{At large energies the interactions become irrelevant and one expects $\phi_{(l,i\nu)}(R)$ to be proportional to $\phi^{\textrm{free}}_{(l,i\nu)}(R)$.}  so that the difference becomes
\begin{empheq}{alignat=7}
	\hat{\zeta}_{\mathcal{O}_l}(s)&\equiv\zeta_{\mathcal{O}_l}(s)-\zeta_{\rm free}(s)&&=\frac{\sin\pi s}{\pi}\int_{\nu_0}^{\infty}d\nu\,\left(\nu^2-\nu_0^2\right)^{-s}\partial_{\nu}\ln\frac{\phi_{(l,i\nu)}(R)}{\phi^{\textrm{free}}_{(l,i\nu)}(R)}\,.
\end{empheq}
This subtraction is further justified by remembering that we are mainly interested in the $R\rightarrow\infty$ limit, where additional divergences related to the IR cutoff  $R$ appear. The integral at $s=0$ is now finite and we can write
\begin{empheq}{alignat=7}
	\hat{\zeta}'_{\mathcal{O}_l}(0)&=-\ln\frac{\phi_{(l,i\nu_0)}(R)}{\phi^{\textrm{free}}_{(l,i\nu_0)}(R)}+\lim_{\nu\rightarrow\infty}\ln\frac{\phi_{(l,i\nu)}(R)}{\phi^{\textrm{free}}_{(l,i\nu)}(R)}\,,
	&\qquad
	\hat\zeta_{\mathcal{O}_l}(0)&=0\,. \label{bosnonsonedimdet}
\end{empheq}
Such a simple expression for the derivative of the zeta-function is valid only because the radial operators $\mathcal{O}_l$ are one-dimensional.  
Notice from \eqref{eq: eigenvalues radial momentum} that $\lambda(i\nu_0)=0$, so the function $\phi_{(l,i\nu_0)}(\rho)$ is the regular solution to the homogeneous equation $\mathcal{O}_l\psi=0$. This equation is typically much easier to solve than the full eigenvalue problem, if not analytically, numerically. The large $\nu$ limit, on the other hand, will be shown to vanish in the bosonic case after a proper normalization. Of course, this is nothing but the Gelfand-Yaglom representation of one-dimensional determinants \cite{Kirsten:2003py,Kirsten:2007ev}. For $d=2$ we still need to sum over Fourier modes. As mentioned above, the sum is divergent at $s=0$, so we are not ready yet. Nonetheless, $\hat{\zeta}'_{\mathcal{O}_l}(0)$ will appear in the final answer.

A similar line of reasoning can be followed for other boundary conditions, even in presence of zero modes, leading to analogous formulas for $\hat{\zeta}_{\mathcal{O}}(s)$ \cite{Kirsten:2007ev,Dunne:2006ct,Kirsten:2003py}. Indeed, with a few modifications, it can also be applied for the fermionic operators \eqref{eq: 2d operator fermions} \cite{Bordag:1998tg,Bordag:2003at}. In this case, since the differential equation is first order, only half of the components of the spinor eigenfunctions can be constrained by the (local) boundary conditions. A standard choice are bag boundary conditions \cite{Vassilevich:2003xt}. Another subtlety is that fermionic operators usually posses negative eigenvalues, leading to an ambiguity in the definition of the zeta-function. This ambiguity can be avoided by considering instead the squared operator, which is second order and is assumed to have a strictly positive spectrum. It is important to emphasize, however, that the eigenvalues of $\mathcal{O}^2$ should already be determined by those of $\mathcal{O}$. In other words, no additional or incompatible boundary conditions should be imposed on the second half of the eigenspinors when dealing with the second order operator. This last statement means that in the countour representation of $\zeta_{\mathcal{O}_l^2}(s)$, it is enough to consider the regular solution to the eigenvalue problem $\mathcal{O}_l\psi=\lambda(\nu)\psi$ and not $\mathcal{O}_l^2\psi=\lambda(\nu)^2\psi$. For convenience we explicitly separate the positive and negative eigenvalue sectors and write
\begin{empheq}{alignat=7}
	\hat\zeta_{\mathcal{O}_l^2}(s)&=\frac{\sin\pi s}{\pi}\int_{\nu_0}^{\infty}d\nu\,\left(\nu^2-\nu_0^2\right)^{-s}\partial_{\nu}\left(\ln\frac{\phi^+_{(l,i\nu)}(R)}{\phi^{+\,\textrm{free}}_{(l,i\nu)}(R)}+\ln\frac{\phi^-_{(l,i\nu)}(R)}{\phi^{-\,\textrm{free}}_{(l,i\nu)}(R)}\right)\,.
\end{empheq}
Here $\phi^{\pm}_{(l,\nu)}(R)$ is some combination, determined by the choice of boundary conditions, of the components of the regular solution to the first order equation $\mathcal{O}_l\psi^{\pm}=\pm\sqrt{\nu^2+\nu_0^2}\,\psi^{\pm}$. The spectrum of the free massive Dirac operator is symmetric, so $\phi^{+\,\textrm{free}}_{(l,\nu)}(R)=\phi^{-\,\textrm{free}}_{(l,\nu)}(R)$, but this is not necessarily the case for interacting operators. Notice the appearance of $\lambda(\nu)^{-2s}$ as opposed to $\lambda(\nu)^{-s}$, meaning that we are squaring the eigenvalues and therefore computing $\hat\zeta_{\mathcal{O}_l^2}(s)$. Evaluating at $s=0$ we get
\begin{empheq}{alignat=7}\label{eq: zetap 1d fermions}
	\hat{\zeta}'_{\mathcal{O}^2_l}(0)&=-\ln\frac{\phi^+_{(l,i\nu_0)}(R)}{\phi^{+\,\textrm{free}}_{(l,i\nu_0)}(R)}-\ln\frac{\phi^-_{(l,i\nu_0)}(R)}{\phi^{-\,\textrm{free}}_{(l,i\nu_0)}(R)}+\lim_{\nu\rightarrow\infty}\left(\ln\frac{\phi^+_{(l,i\nu)}(R)}{\phi^{+\,\textrm{free}}_{(l,i\nu)}(R)}+\ln\frac{\phi^-_{(l,i\nu)}(R)}{\phi^{-\,\textrm{free}}_{(l,i\nu)}(R)}\right)\,,
	&\qquad
	\hat\zeta_{\mathcal{O}_l^2}(0)&=0\,.
\end{empheq}
Again, the computation of the zeta-function for the full fermionic operator requires a summation over the (half-integer) Fourier modes, so we are not allowed to take $s=0$ at this moment.
\subsection{Free eigenfunctions, Jost function and boundary conditions}
We are interested in operators of the form \eqref{eq: 2d operator bosons} and \eqref{eq: 2d operator fermions} for which the background fields decay sufficiently fast at infinity, so that they become effectively free. Therefore, it is not surprising that the free eigenfunctions play a preponderant role in the analysis. Their exact form will be displayed below. For the moment we focus on some of their properties. 

Let $h_{\pm}^{(l,\nu)}(\rho)$ be the two linearly independent eigenfunctions of the operator $\mathcal{O}_l^{\textrm{free}}$. They satisfy
\begin{empheq}{alignat=7}\label{eq: free eigenfunctions equation}
	\mathcal{O}_l^{\textrm{free}}h_{\pm}^{(l,\nu)}(\rho)&=\lambda(\nu)h_{\pm}^{(l,\nu)}(\rho)\,,
\end{empheq}
where the eigenvalues are parametrized as in \eqref{eq: eigenvalues radial momentum}. In the fermionic case these are actually two-component eigenspinors and should carry an additional label specifying the sign of the eigenvalues. The notation $\pm$ refers to the fact that, asymptotically, these solutions become in- and out-going waves,
\begin{empheq}{alignat=7}
	h_{\pm}^{(l,\nu)}(\rho)&\sim e^{\left(-\frac{1}{2}\pm i\nu\right)\rho}\,,
	&\qquad
	\rho&\rightarrow\infty\,,
\end{empheq}
as follows directly from the differential equation. Square-integrability requires that $\nu\in\mathds{R}$; the modulating factor $e^{-\frac{\rho}{2}}$ is compensated by the integration measure $\sqrt{g}=\sinh\rho\sim e^{\rho}$, yielding a plane wave orthogonality relation. It is important to mention, however, that neither $h_+^{(l,\nu)}(\rho)$ nor $h_-^{(l,\nu)}(\rho)$ are regular at $\rho=0$, and therefore not actually square-integrable. Rather, after an appropriate choice of relative normalizations, the free regular solution is given by the combination
\begin{empheq}{alignat=7}
	\phi_{(l,\nu)}^{\textrm{free}}(\rho)&=\frac{i}{2}\left(h_-^{(l,\nu)}(\rho)-h_+^{(l,\nu)}(\rho)\right)\,.
\end{empheq}
Its small $\rho$ expansion is again dictated by the differential equation and reads
\begin{empheq}{alignat=7}
	\phi^{\textrm{free}}_{(l,\nu)}(\rho)&\sim\rho^{|l|}\,,
	&\qquad
	\rho&\rightarrow0\,,
	&\qquad\textrm{(bosons)}
	\\
	\phi^{\textrm{free}}_{(l,\nu)}(\rho)&\sim\rho^{|l|-\frac{1}{2}}\,,
	&\qquad
	\rho&\rightarrow0\,,
	&\qquad\textrm{(fermions)}
\end{empheq}
For fermions only for the leading component is shown; the other component goes like $\rho^{|l|+\frac{1}{2}}$. The overall constant will depend on the exact normalization of $h_{\pm}^{(l,\nu)}$, the choice of which is arbitrary.

Consider now the interacting case. In general, finding the regular solution is prohibitively complicated. Nevertheless, there are two statements that are generally true. The first is that, precisely because it is regular, the behavior of $\phi_{(l,\nu)}(\rho)$ at $\rho=0$ is the same as for the free solution. The second property stems from the previous observation that the operators become free for large $\rho$, meaning that the regular solution can be expanded as\footnote{Given that the gauge field goes to a constant  $\mathcal{A}(\rho)\rightarrow \mathcal{A}_{\infty}$ for $\rho\rightarrow\infty$, the asymptotics of the regular solution is more naturally expanded in terms of the shifted eigenfunctions $h^{(l-\mathcal{A}_{\infty},\nu)}_{\pm}(\rho)$. At large $\rho$, however, these differ from their un-shifted version only by a normalization, making the definition \eqref{jostdefinition} of the Jost function still viable.}
\begin{empheq}{alignat=7}
	\phi_{(l,\nu)}(\rho)&\underset{\rho\rightarrow\infty}{\longrightarrow}\frac{i}{2}\left(g_l(\nu)h_-^{(l,\nu)}(\rho)-\bar{g}_l(\nu)h_+^{(l,\nu)}(\rho)\right)\,.
\label{jostdefinition}
\end{empheq}
This is only true asymptotically, of course. The coefficient $g_l(\nu)$ is called Jost function and  plays a central part in the calculation of functional determinants. In fact, the the ratio $\ln\left(g_l(\nu)/\bar{g}_l(\nu)\right)$ is precisely the phase shift from scattering theory that determines the density of eigenvalues. In the free case the above relation becomes exact with $g_l^{\textrm{free}}(\nu)=1$.

Let us use the properties we have just discussed to see what happens to the zeta-function when we take the infinite space limit $R\rightarrow\infty$. To this purpose, note that for imaginary values of the radial momentum, the function $h_+^{(l,i\nu)}(R)$ is exponentially decaying, whereas $h_-^{(l,i\nu)}(R)$ blows up. Therefore, the ratio between the regular interacting solution and the free one becomes
\begin{empheq}{alignat=7}
	\lim_{R\rightarrow\infty}\frac{\phi_{(l,i\nu)}(R)}{\phi^{\textrm{free}}_{(l,i\nu)}(R)}&=g_l(i\nu)\,.
\label{Jostsolution}
\end{empheq}
This gives the following expression for the zeta-function of the bosonic operator \eqref{eq: 2d operator bosons}
\begin{empheq}{alignat=7}\label{eq: zeta function bosons Jost}
	\hat{\zeta}_{\mathcal{O}}(s)&=\frac{\sin\pi s}{\pi}\sum_{l\,\in\,\mathds{Z}}\int_{\nu_0}^{\infty}d\nu\,\left(\nu^2-\nu_0^2\right)^{-s}\partial_{\nu}\ln g_l(i\nu)\,.
\end{empheq}
A similar simplification occurs in the fermionic case \eqref{eq: 2d operator fermions}, yielding
\begin{empheq}{alignat=7}\label{eq: zeta function fermions Jost}
	\hat{\zeta}_{\mathcal{O}^2}(s)&=\frac{\sin\pi s}{\pi}\sum_{l\,\in\,\mathds{Z}+\frac{1}{2}}\int_{\nu_0}^{\infty}d\nu\,\left(\nu^2-\nu_0^2\right)^{-s}\partial_{\nu}\ln {\mathfrak g}_l(i\nu)\,,
\end{empheq}
where $\ln \mathfrak{g}_l(i\nu) \equiv \ln g^+_l(i\nu)+\ln g^-_{l}(i\nu) $ includes the contribution from the positive and negative eigenvalue sectors. Technically, the above expressions define the zeta function  in terms of scattering data.

Besides the introduction of the Jost function in the two formulas above, the $R\rightarrow\infty$ limit has another, crucial, consequence on the zeta function: it makes the dependence on the specific choice of boundary conditions disappear. Take for example the case of Neumann boundary conditions. The only modification one needs to make in $\hat{\zeta}_{\mathcal{O}}(s)$ is the replacement $\phi_{(l,i\nu)}(R)\rightarrow\partial_{\rho}\phi_{(l,i\nu)}(R)$. It is easy to see that upon taking the ratio with the corresponding free solution, the large $R$ limit will again be given by the Jost function. The same is true for more general boundary conditions and for spinor fields. We then conclude that the determinants in $AdS_2$ are insensitive to the choice of boundary conditions one makes in the intermediate step of putting the system in a finite box.

As pointed out several times already, the sum over Fourier modes is ill-defined for $s=0$. In what follows, we will perform the analytic continuation of \eqref{eq: zeta function bosons Jost} and \eqref{eq: zeta function fermions Jost}. The general strategy is to subtract as many terms as necessary inside the integral such that the series becomes convergent at $s=0$. The dangerous region is obviously $l\rightarrow\infty$, but also $\nu\sim l\rightarrow\infty$, so the calculation involves extracting the asymptotic behavior of $g_l(i\nu)$ in this regime. This can be done by constructing a representation of the Jost function in terms of the free eigenfunctions $h_{\pm}^{(l,\nu)}(\rho)$, the Green's function for the free operator and the background fields. The subtracted terms need to be added back and the analytic continuation is done using the well-known properties of the Riemann zeta-function.


\subsection{Bosons}
In this section we exhibit the derivation of \eqref{eq: main result bosons}. We split the radial operator \eqref{eq: 1d radial operator bosons} into a free part and an interaction,
\begin{empheq}{alignat=7}
	\mathcal{O}_l&=\mathcal{O}_l^{\textrm{free}}+U(\rho)\,,
	&\qquad
	U(\rho)&=V(\rho)+\frac{\mathcal{A}(\rho)^2}{\sinh^2\rho}-\frac{2l\mathcal{A}(\rho)}{\sinh^2\rho}\,.
\label{bosonic potential}
\end{empheq}
where the free operator is given by
\begin{empheq}{alignat=7}
	\mathcal{O}_l^{\textrm{free}}&=-\frac{1}{\sinh\rho}\partial_{\rho}\left(\sinh\rho\,\partial_{\rho}\right)+\frac{l^2}{\sinh^2\rho}+m^2\,,
	&\qquad
	l&\in\mathds{Z}\,.
\end{empheq}
It will be important in what follows to keep in mind that there is a $ l $-dependent term in the potential $U(\rho)$.


\subsubsection{Free eigenfunctions}
The bosonic free eigenfunctions satisfying \eqref{eq: free eigenfunctions equation} read
\begin{empheq}{alignat=7}
	h^{(l,\nu)}_{\pm}(\rho)&=\sqrt{\frac{2}{\pi\nu}}\left|\frac{\Gamma\left(1+i\nu\right)}{\Gamma\left(\frac{1}{2}+i\nu+|l|\right)}\right|e^{-i\pi|l|}Q^{|l|}_{-\frac{1}{2}\mp i\nu}(\cosh\rho)\,,
	&\qquad
	\left(h^{(l,\nu)}_{\pm}\right)^*&=h^{(l,\nu)}_{\mp}\,,
\end{empheq}
where $Q^{|l|}_{-\frac{1}{2}\mp i\nu}(\cosh\rho)$ are associated Legendre functions of the second kind. The condition that $\nu\in\mathds{R}$ is necessary for square-integrability, as can be seen from the asymptotic expansions
\begin{empheq}{alignat=7}\label{eq: asymptotic expansions}
\begin{split}
	h^{(l,\nu)}_{\pm}(\rho)&\approx\sqrt{\frac{2}{\nu}}\left|\frac{\Gamma\left(1+i\nu\right)}{\Gamma\left(\frac{1}{2}+i\nu+|l|\right)}\right|\frac{\Gamma\left(\frac{1}{2}\mp i\nu+|l|\right)}{\Gamma\left(1\mp i\nu\right)}e^{\left(-\frac{1}{2}\pm i\nu\right)\rho}\,,
\end{split}
\qquad
\begin{split}
	\rho&\rightarrow\infty\,.
\end{split}
\end{empheq}
The combination
\begin{empheq}{alignat=7}
	\phi^{\rm free}_{(l,\nu)}(\rho)&\equiv\frac{i}{2}\left(h^{(l,\nu)}_-(\rho)-h^{(l,\nu)}_+(\rho)\right)&&=\sqrt{\frac{\pi\nu}{2}}\left|\frac{\Gamma\left(\frac{1}{2}+i\nu+|l|\right)}{\Gamma\left(1+i\nu\right)}\right|P_{-\frac{1}{2}\pm i\nu}^{-|l|}\left(\cosh\rho\right)\,,
\end{empheq}
namely, the imaginary part of the eigenfunctions, is proportional to the associated Legendre function of the first kind and is regular at $\rho=0$ with
\begin{empheq}{alignat=7}
	\phi^{\rm free}_{(l,\nu)}(\rho)&\approx\sqrt{\frac{\pi\nu}{2}}\left|\frac{\Gamma\left(\frac{1}{2}+i\nu+|l|\right)}{\Gamma\left(1+i\nu\right)}\right|\frac{1}{\Gamma\left(1+|l|\right)}\left(\frac{\rho}{2}\right)^{|l|}\,,
	&\qquad
	\rho&\rightarrow0\,.
\end{empheq}
As a matter of convenience, the normalization of the eigenfunctions has been chosen so that their Wronskian is independent of $\nu$:
\begin{empheq}{alignat=7}\label{eq: Wronskian}
	h^{(l,\nu)}_-(\rho)\partial_{\rho}h^{(l,\nu)}_+(\rho)-h^{(l,\nu)}_+(\rho)\partial_{\rho}h^{(l,\nu)}_-(\rho)&=\frac{2i}{\sinh\rho}\,.
\end{empheq}
Regardless of the normalization, this property allows us to construct the Green's function
\begin{empheq}{alignat=7}\label{eq: Green's function bosons}
	G^{(l,\nu)}(\rho,\rho')&=\frac{i}{2}\sinh\rho'\left(h^{(l,\nu)}_-(\rho)h^{(l,\nu)}_+(\rho')-h^{(l,\nu)}_+(\rho)h^{(l,\nu)}_-(\rho')\right)\theta(\rho-\rho')
	\\
	&=\sinh\rho'\left(\phi^{\rm free}_{(l,\nu)}(\rho)h^{(l,\nu)}_+(\rho')-\phi^{\rm free}_{(l,\nu)}(\rho)h^{(l,\nu)}_-(\rho')\right)\theta(\rho-\rho')\,,\nn
\end{empheq}
which satisfies
\begin{empheq}{alignat=7}
	\left(\mathcal{O}_l^{\textrm{free}}-\lambda(\nu)\right)G^{(l,\nu)}(\rho,\rho')&=-\delta(\rho,\rho')\,.
\end{empheq}
Finally, we need to continue the eigenfunctions to imaginary momentum, $\nu\rightarrow i\nu$, and extract their asymptotic behavior for $l\rightarrow\infty$ and fixed $ \alpha\equiv\frac{\nu}{|l|}$ with $0<\alpha<1$. We find
\begin{empheq}{alignat=7}\label{hasymptbosons}
	h^{(l,i\nu)}_+(\rho)&\approx\sqrt{\frac{\alpha}{\pi|\sin\left(\pi\nu\right)|}}\left(1-\alpha^2\right)^{\frac{\nu+|l|}{2}}\left(\alpha^2\sinh^2\rho+1\right)^{-\frac{1}{4}}e^{-|l| \eta(\rho) }\,,
	\\
	\phi^{\rm free}_{(l,i\nu)}(\rho)&\approx i\sqrt{\frac{\alpha|\sin\left(\pi\nu\right)|}{\pi}}\left(1-\alpha^2\right)^{-\frac{\nu+|l|}{2}}\left(\alpha^2\sinh^2\rho+1\right)^{-\frac{1}{4}}e^{|l| \eta(\rho) }\,,\label{jasymptbosons}
\end{empheq}
where
\be
\eta(\rho) = \alpha\ln\left(\alpha\cosh\rho+\sqrt{1+\alpha^2\sinh^2\rho}\right)- \ln\left(\cosh\rho+\sqrt{1+\alpha^2\sinh^2\rho}\right)+\ln\sinh\rho\, .
\label{etarho}
\ee


\subsubsection{Regular solution and Jost function}
In order to compute the zeta-function using \eqref{eq: zeta function bosons Jost}, we first need to construct a solution to the eigenvalue problem that is regular at the origin. With the help of the free Green's function \eqref{eq: Green's function bosons}, we can invert the differential equation and write it in Lippmann-Schwinger form,
\begin{empheq}{alignat=7}\label{eq: integral equation bosons}
	\phi_{(l,\nu)}(\rho)&=\phi^{\rm free}_{(l,\nu)}(\rho)+\int_0^{\rho}d\rho' G^{(l,\nu)}(\rho,\rho')U(\rho')\phi_{(l,\nu)}(\rho')\,.
\end{empheq}
In principle the integral above extends to $\rho'\rightarrow\infty$, but our choice of Green's function truncates it to $\rho'\le \rho$. This choice is dictated by the fact that we want to control the behavior of the solution at $\rho=0$ to ensure that it is regular. Notice that $G^{(l,\nu)}(\rho,\rho)=0$, so the normalization $\phi_{(l,\nu)}(\rho)\approx\phi^{\rm free}_{(l,\nu)}(\rho)$, with the same leading coefficient in the series expansion, is fixed by the integral equation.

Replacing the Green's function \eqref{eq: Green's function bosons} in \eqref{eq: integral equation bosons}, taking $\rho\rightarrow\infty$ and by means of  \eqref{jostdefinition}, we arrive to the following expression for the Jost function
\begin{empheq}{alignat=7}
	g_l(\nu)&=1+\int_0^{\infty}d\rho\,\sinh\rho\,h_+^{(l,\nu)}(\rho)U(\rho)\phi_{(l,\nu)}(\rho)\,.
\end{empheq}
Of course, this expression still involves the unknown function $\phi^{(l,\nu)}(\rho)$ and can be solved iteratively as an expansion in powers of the potential $U$. However, as we will confirm below, it is sufficient to solve for the regular solution only up to second order. After some algebra one gets\footnote{Use ${\displaystyle\ln\left(1+ax+bx^2\right)=ax+\left(b-\frac{1}{2}a^2\right)x^2+O(x^3)}$ and ${\displaystyle\int_a^bdxf(x)\int_a^xdyf(y)=\frac{1}{2}\left(\int_a^bdxf(x)\right)^2}$.}
\begin{empheq}{alignat=7}
	\ln g_l(\nu)=&\int_0^{\infty}d\rho\,\sinh\rho\,h_+^{(l,\nu)}(\rho)U(\rho)\phi^{\rm free}_{(l,\nu)}(\rho)
	\label{perturbativejostbosons}\\
	&-\int_0^{\infty}d\rho\,\sinh\rho\,\left(h_+^{(l,\nu)}(\rho)\right)^2U(\rho)\int_0^{\rho}d\rho'\,\sinh\rho'\,\left(\phi^{\rm free}_{(l,\nu)}(\rho')\right)^2 U(\rho')+O(U^3)\,,\nn
\end{empheq}
where we have taken the logarithm since that is what actually enters in the $\zeta$-function.

The next step involves continuing the Jost function to imaginary values of the radial momentum and extracting its limiting behavior for large $\nu$ and large $l$. Remember that the goal is to subtract from $\ln g_l(i\nu)$ as many terms as necessary so that the sum over Fourier modes in \eqref{eq: zeta function bosons Jost} becomes convergent at $s=0$. Clearly we can discard all terms that decay faster than $l^{-1}$. Introducing the asymptotic expansions of the eigenfunctions given in \eqref{hasymptbosons} and \eqref{jasymptbosons} into \eqref{perturbativejostbosons} we obtain
\begin{empheq}{alignat=7}\label{gasymbosons}
	\ln g_l(i\nu)&=\frac{1}{2|l|}\int_0^{\infty}d\rho\frac{\sinh\rho\, U(\rho)}{\sqrt{\alpha^2\sinh^2\rho+1}}
	\\\nonumber
	&-\frac{1}{4l^2}\int_0^{\infty}d\rho\,\frac{\sinh\rho\,U(\rho)e^{-2|l| \eta(\rho)}}{\sqrt{\alpha^2\sinh^2\rho+1}}
\int_0^{\rho}d\rho'\,\frac{\sinh\rho' U(\rho')e^{2|l| \eta(\rho')}}{\sqrt{\alpha^2\sinh^2\rho'+1}}+O(l^{-2})\,.
\end{empheq}
Notice that the first line involves a term of order $O(l^0)$ coming from \eqref{bosonic potential}. However, this will cancel when summing over positive and negative Fourier modes. By the same token, subleading contributions to eigenfunctions where not considered in \eqref{hasymptbosons} and \eqref{jasymptbosons}, as they are insensitive to the sign of $l$. A priori, the second line also involves a $O(l^0)$ term, but this is really not so. It can be seen that in the saddle point approximation, which is justified in the limit we are studying, the integral over $\rho'$ yields
\begin{empheq}{alignat=7}
		\int_0^{\rho}d\rho'\,\frac{\sinh\rho' U(\rho')e^{2|l| \eta(\rho')}}{\sqrt{\alpha^2\sinh^2\rho'+1}}\approx\frac{1}{2|l|}\frac{\sinh^2\rho\,U(\rho)e^{2|l| \eta(\rho)}}{\alpha^2\sinh^2\rho+1}+O({l}^{-2})\,.
\end{empheq} 
Since each nested integral results in a factor of $1/l$, higher orders in $U$ in the Lippmann-Schwinger  expansion \eqref{eq: integral equation bosons} are not necessary for the subtraction. This way we arrive at the following expression for the asymptotic behavior of the Jost function 
\begin{empheq}{alignat=7}\label{eq: Jost function asymptotic bosons}
	\ln g^{\textrm{asym}}_l(i\nu)+\ln g^{\textrm{asym}}_{-l}(i\nu)&\equiv\frac{1}{|l|}\int_0^{\infty}d\rho\,\frac{\sinh\rho\,V(\rho)}{\left(1+\alpha^2\sinh^2\rho\right)^{\frac{1}{2}}}+\frac{\alpha^2}{|l|}\int_0^{\infty}d\rho\,\frac{\sinh\rho\,\mathcal{A}(\rho)^2}{\left(1+\alpha^2\sinh^2\rho\right)^{\frac{3}{2}}}\,.
\end{empheq}
Recall that the dependence on the radial momentum enters through $\alpha=\nu/|l|$. One can easily see that
\begin{empheq}{alignat=7}
	\lim_{\nu\rightarrow\infty}\left(\ln g^{\textrm{asym}}_l(i\nu)+\ln g^{\textrm{asym}}_{-l}(i\nu)\right)&=0\,.
\end{empheq}
Similarly, expanding \ref{perturbativejostbosons} for large $\nu$ and fixed $l$ one finds\footnote{We omit the explicit expansions of the eigenfunctions in this limit since they are even simpler than the ones presented above. }
\begin{empheq}{alignat=7}
	\lim_{\nu\rightarrow\infty}\ln g_l(i\nu)&=0\,.\label{bosonsginfty}
\end{empheq}
The fact that this limit vanishes is a consequence of the choice of normalization of the regular solution.
\subsubsection{Analytic continuation}
The analytic continuation of the zeta-function \eqref{eq: zeta function bosons Jost} to $s=0$ is achieved by splitting the sum as
\begin{empheq}{alignat=7}
	\hat{\zeta}_{\mathcal{O}}(s)&=\hat{\zeta}_f(s)+\hat{\zeta}_d(s)\,,
\end{empheq}
where
\begin{empheq}{alignat=7}
	\hat{\zeta}_f(s)=&\frac{\sin\pi s}{\pi}\int_{\nu_0}^{\infty}d\nu\left(\nu^2-\nu_0^2\right)^{-s}\partial_{\nu}\ln g_0(i\nu)
	\\
	&+\frac{\sin\pi s}{\pi}\sum_{l=1}^{\infty}\int_{\nu_0}^{\infty}d\nu\left(\nu^2-\nu_0^2\right)^{-s}\partial_{\nu}\left(\ln g_l(i\nu)+\ln g_{-l}(i\nu)-\ln g^{\textrm{asym}}_l(i\nu)-\ln g^{\textrm{asym}}_{-l}(i\nu)\right)\,,\nn
	\\
	\hat{\zeta}_d(s)=&\frac{\sin\pi s}{\pi}\sum_{l=1}^{\infty}\int_{\nu_0}^{\infty}d\nu\left(\nu^2-\nu_0^2\right)^{-s}\partial_{\nu}\left(\ln g^{\textrm{asym}}_l(i\nu)+\ln g^{\textrm{asym}}_{-l}(i\nu)\right)\,.
\end{empheq}
Here we have separated the mode $l=0$ and combined the $l>0$ and $l<0$ terms into a single sum. The main point is that $\zeta_f(s)$ is now convergent at $s=0$, since by construction of $g^{\textrm{asym}}_l(i\nu)$ it goes as $O(l^{-2})$ for $l\rightarrow\infty$. Thus, we can simply take its derivative and evaluate
\begin{empheq}{alignat=7}
	\hat{\zeta}_f(0)&=0\,,
	\\
	\hat{\zeta}'_f(0)&=-\ln g_0\left(i\nu_0\right)-\sum_{l=1}^{\infty}\left(\ln g_l\left(i\nu_0\right)+\ln g_{-l}\left(i\nu_0\right)-\ln g_l^{\textrm{asym}}\left(i\nu_0\right)-\ln g_{-l}^{\textrm{asym}}\left(i\nu_0\right)\right)\,.
\end{empheq}
Again, $\hat{\zeta}'_f(0)$ is guaranteed to be finite. On the other hand, $\zeta_d(s)$ is still divergent at $s=0$ and needs continuation. The improvement is that this sum is easier to handle. Indeed, the general formulas
\begin{empheq}{alignat=7}
	\int_a^{\infty}dx\left(x^2-a^2\right)^{-s}\frac{d}{dx}\left(\left(1+b^2x^2\right)^{-n/2}\right)&=-\frac{\Gamma\left(s+\frac{n}{2}\right)\Gamma\left(1-s\right)b^{2s}}{\Gamma\left(\frac{n}{2}\right)\left(1+a^2b^2\right)^{s+\frac{n}{2}}}\,,
	\label{intform1}\\
	\int_a^{\infty}dx\left(x^2-a^2\right)^{-s}\frac{d}{dx}\left(x^2\left(1+b^2x^2\right)^{-n/2}\right)&=-\frac{\Gamma\left(s+\frac{n}{2}-1\right)\Gamma\left(1-s\right)b^{2(s-1)}\left(\left(n-2\right)a^2b^2-2s\right)}{2\Gamma\left(\frac{n}{2}\right)\left(1+a^2b^2\right)^{s+\frac{n}{2}}}\,,\label{intform2}
\end{empheq}
allow us to explicitly perform the integration over the radial momentum and find
\begin{empheq}{alignat=7}
	\zeta_d(s)&=-\frac{\Gamma\left(s+\frac{1}{2}\right)\Gamma\left(1-s\right)}{\Gamma\left(\frac{1}{2}\right)}\int_0^{\infty}d\rho\,\left(\sinh\rho\right)^{2s+1}\left(V(\rho)R_1(s,\rho)+\frac{\mathcal{A}(\rho)^2}{\sinh^2\rho}R_2(s,\rho)\right)\,,
\end{empheq}
where
\begin{empheq}{alignat=7}
	R_1(s,\rho)&=\frac{\sin\pi s}{\pi}\sum_{l=1}^{\infty}\frac{1}{l^{1+2s}}\left(1+\frac{\nu_0^2\sinh^2\rho}{l^2}\right)^{-\left(s+\frac{1}{2}\right)}\,,\label{R1}
	\\
	R_2(s,\rho)&=\frac{\sin\pi s}{\pi}\sum_{l=1}^{\infty}\frac{1}{l^{1+2s}}\left(\frac{\nu_0^2\sinh^2\rho}{l^2}-2s\right)\left(1+\frac{\nu_0^2\sinh^2\rho}{l^2}\right)^{-\left(s+\frac{3}{2}\right)}\,.\label{R2}
\end{empheq}
In order to continue these sums, we again subtract and add back the asymptotic behavior of the summand that makes the series divergent when $s=0$, namely,
\begin{empheq}{alignat=7}
	R_1(s,\rho)&=\frac{\sin\pi s}{\pi}\sum_{l=1}^{\infty}\frac{1}{l^{1+2s}}\left[\left(1+\frac{\nu_0^2\sinh^2\rho}{l^2}\right)^{-\left(s+\frac{1}{2}\right)}-1\right]+\frac{\sin\pi s}{\pi}\sum_{l=1}^{\infty}\frac{1}{l^{1+2s}}\,,\nn
	\\
	R_2(s,\rho)&=\frac{\sin\pi s}{\pi}\sum_{l=1}^{\infty}\frac{1}{l^{1+2s}}\left[\left(\frac{\nu_0^2\sinh^2\rho}{l^2}-2s\right)\left(1+\frac{\nu_0^2\sinh^2\rho}{l^2}\right)^{-\left(s+\frac{3}{2}\right)}+2s\right]-\frac{2s\sin\pi s}{\pi}\sum_{l=1}^{\infty}\frac{1}{l^{1+2s}}\,.\nn
\end{empheq}
Recognizing the last term in each expression as the Riemann zeta function, we arrive at
\begin{empheq}{alignat=7}
	R_1(s,\rho)=&\frac{\sin\pi s}{\pi}\sum_{l=1}^{\infty}\frac{1}{l^{1+2s}}\left[\left(1+\frac{\nu_0^2\sinh^2\rho}{l^2}\right)^{-\left(s+\frac{1}{2}\right)}-1\right]+\frac{\sin\pi s}{\pi}\zeta_R(2s+1)\,,
	\\
	R_2(s,\rho)=&\frac{\sin\pi s}{\pi}\sum_{l=1}^{\infty}\frac{1}{l^{1+2s}}\left[\left(\frac{\nu_0^2\sinh^2\rho}{l^2}-2s\right)\left(1+\frac{\nu_0^2\sinh^2\rho}{l^2}\right)^{-\left(s+\frac{3}{2}\right)}+2s\right]
	\\
	&-\frac{2s\sin\pi s}{\pi}\zeta_R(2s+1)\,.\nn
\end{empheq}
Since each sum in square brackets is now convergent for $s=0$, we readily find\footnote{Actually, $R_2(s,\rho)$ was already convergent at $s=0$. However, its term by term derivative was not, so the procedure was still necessary.}
\begin{empheq}{alignat=7}
	R_1(0,\rho)&=\frac{1}{2}\,,
	&\qquad
	R_1'(0,\rho)&=\sum_{l=1}^{\infty}\frac{1}{l}\left[\left(1+\frac{\nu_0^2\sinh^2\rho}{l^2}\right)^{-\frac{1}{2}}-1\right]+\gamma\,,
	\\
	R_2(0,\rho)&=0\,,	
	&\qquad
	R_2'(0,\rho)&=\nu_0^2\sinh^2\rho\sum_{l=1}^{\infty}\frac{1}{l^3}\left(1+\frac{\nu_0^2\sinh^2\rho}{l^2}\right)^{-\frac{3}{2}}-1\,.
\end{empheq}
This is the desired continuation. Then,
\begin{empheq}{alignat=7}
	\hat{\zeta}_d(0)&=-\frac{1}{2}\int_0^{\infty}d\rho\,\sinh\rho\,V(\rho)\,,
\end{empheq}
and
\begin{empheq}{alignat=7}
	\hat{\zeta}'_d(0)&=-\int_0^{\infty}d\rho\,\sinh\rho\left(\ln\left(\frac{\sinh\rho}{2}\right)+\gamma\right)V(\rho)+\int_0^{\infty}d\rho\,\frac{\mathcal{A}(\rho)^2}{\sinh\rho}\nn
	\\
	&\quad -\sum_{l=1}^{\infty}\frac{1}{l}\int_0^{\infty}d\rho\,\sinh\rho\left[\left(1+\frac{\nu_0^2\sinh^2\rho}{l^2}\right)^{-\frac{1}{2}}V(\rho)-V(\rho)-\frac{\nu_0^2}{l^2}\left(1+\frac{\nu_0^2\sinh^2\rho}{l^2}\right)^{-1}\mathcal{A}(\rho)^2\right]\nn
	\\
	&=-\int_0^{\infty}d\rho\,\sinh\rho\left(\ln\left(\frac{\sinh\rho}{2}\right)+\gamma\right)V(\rho)+\int_0^{\infty}d\rho\,\frac{\mathcal{A}(\rho)^2}{\sinh\rho}
	\\
	&\quad -\sum_{l=1}^{\infty}\left(\ln g^{\textrm{asym}}_{l}(i\nu_0)+\ln g^{\textrm{asym}}_{-l}(i\nu_0)-\frac{1}{l}\int_0^{\infty}d\rho\,\sinh\rho\,V(\rho)\right)\,.\nn
\end{empheq}
In the last step we have recognized the asymptotic form \eqref{eq: Jost function asymptotic bosons} of the Jost function evaluated at $\nu=\nu_0$. Combining the expressions for $\hat{\zeta}_f(0)$, $\hat{\zeta}_d(0)$, $\hat{\zeta}'_f(0)$ and $\hat{\zeta}'_d(0)$ we arrive at
\begin{empheq}{alignat=7}
	\hat{\zeta}_{\mathcal{O}}(0)&=-\frac{1}{2}\int_0^{\infty}d\rho\,\sinh\rho\,V(\rho)\,,
	\\
	\hat{\zeta}'_{\mathcal{O}}(0)&=-\ln g_0\left(i\nu_0\right)-\sum_{l=1}^{\infty}\left(\ln g_l\left(i\nu_0\right)+\ln g_{-l}\left(i\nu_0\right)+\frac{2}{l}\hat{\zeta}(0)\right)+2\gamma\hat{\zeta}(0)
	\\
	& \quad -\int_0^{\infty}d\rho\,\sinh\rho\ln\left(\frac{\sinh\rho}{2}\right)V(\rho)+\int_0^{\infty}d\rho\,\frac{\mathcal{A}(\rho)^2}{\sinh\rho}\,.\nn
\end{empheq}
Notice that $\ln g^{\textrm{asym}}_{l}(i\nu_0)$ cancels out at the end so it is no longer needed. Finally, by means of \eqref{Jostsolution}, \eqref{bosnonsonedimdet} and \eqref{bosonsginfty}, $g_l(i\nu_0)$ is identified with the determinant of the radial operator $\mathcal{O}_l$ and the full renormalized determinant \eqref{eq: renormalized determinant} becomes our main result \eqref{eq: main result bosons}. Once the radius of $AdS_2$ is reinstated, the dimensionless quantity $L\mu$ appears.


 \subsection{Fermions}

We now move on to the derivation of the fermionic expression \eqref{eq: main result fermions}. As in the bosonic case, the full operator splits into
\begin{empheq}{alignat=7}
	\mathcal{O}_l&=\mathcal{O}^{\textrm{free}}_l-i\Gamma_{\underline{01}}U(\rho)
	&\quad , \quad
	U(\rho)&=-\Gamma_{\underline{0}}\,\partial_{\rho}\Omega(\rho)-i \, q\, \Gamma_{\underline{1}}\frac{\mathcal{A}(\rho)}{\sinh\rho}+V(\rho)-i\Gamma_{\underline{01}}W(\rho)\,. \label{interacting fermion}
\end{empheq}
The matrix $-i\Gamma_{\underline{01}}$ in front of $U$ is a matter of convenience.
The free fermionic radial operator is
\begin{empheq}{alignat=7}
	\mathcal{O}^{\textrm{free}}_l&=-i\Gamma_{\underline{1}}\left(\partial_{\rho}+\frac{1}{2}\coth\rho\right)+\Gamma_{\underline{0}}\frac{l}{\sinh\rho}-i\Gamma_{\underline{01}}m\,,
	&\qquad
	l&\in\mathds{Z}+\frac{1}{2}\,. \label{free fermion}
\end{empheq}
From now on we will work with the following representation of the Dirac matrices,
\be 
\Gamma_{\underline{0}} = -\sigma_2\,, \quad \Gamma_{\underline{0}} = \sigma_1 \quad \Rightarrow \quad -i\Gamma_{\underline{01}} = \sigma_3\,.
\ee

\subsubsection{Free eigenfunctions}

Unlike the bosonic case, the free operator \eqref{free fermion} has positive and negative eigenvalues. It is sufficient, however, to restrict ourselves to $\lambda>0$, since the $\lambda<0$ sector can be obtained from the former by a simple operation. The eigenfunctions for $l\geq\frac{1}{2}$ and $l\leq-\frac{1}{2}$ are also related to each other, so we will work with strictly positive Fourier modes. This is not to say that we are              neglecting three out of the four possible sectors.

The spinor eigenfunctions satisfying \eqref{eq: free eigenfunctions equation} with $\lambda>0$ and $l\geq\frac{1}{2}$ read
\begin{empheq}{alignat=7}\label{eq: free eigenfunctions fermions}
	h_{\pm}^{(l,\nu)}(\rho)&=\sqrt{\frac{\Gamma\left(l+\frac{1}{2}\mp i\nu\right)\Gamma\left(\frac{1}{2}\pm i\nu\right)}{\Gamma\left(l+\frac{1}{2}\pm i\nu\right)\Gamma\left(\frac{1}{2}\mp i\nu\right)}}\sqrt{2}\left(\tanh\frac{\rho}{2}\right)^{l-\frac{1}{2}}\left(2\cosh\frac{\rho}{2}\right)^{-1\pm2i\nu}\psi_{\pm}^{(l,\nu)}(\rho)\,,
\end{empheq}
where
\begin{empheq}{alignat=7}
	\psi_{\pm}^{(l,\nu)}(\rho)&=
	\left(
	\begin{array}{c}
		{\displaystyle\left(\frac{\lambda(\nu)+m}{\lambda(\nu)-m}\right)^{\frac{1}{4}}\tanh\frac{\rho}{2}F\left(l+\frac{1}{2}\mp i\nu,1\mp i\nu;1\mp2i\nu;\frac{1}{\cosh^2\frac{\rho}{2}}\right)}
		\\
		{\displaystyle\pm\left(\frac{\lambda(\nu)-m}{\lambda(\nu)+m}\right)^{\frac{1}{4}}F\left(l+\frac{1}{2}\mp i\nu,\mp i\nu;1\mp2i\nu;\frac{1}{\cosh^2\frac{\rho}{2}}\right)}
	\end{array}
	\right)\,.
\end{empheq}
The combination
\begin{empheq}{alignat=7}
	\phi^{\rm free}_{(l,\nu)}(\rho)&\equiv\frac{i}{2}\left(h_-^{(l,\nu)}(\rho)-h_+^{(l,\nu)}(\rho)\right)\nn
	\\
	&=\frac{1}{\Gamma\left(l+\frac{1}{2}\right)}\sqrt{\frac{\pi}{2}}\left|\frac{\Gamma\left(l+\frac{1}{2}\mp i\nu\right)}{\Gamma\left(\frac{1}{2}\mp i\nu\right)}\right|\left(\tanh\frac{\rho}{2}\right)^{l-\frac{1}{2}}\left(\cosh\frac{\rho}{2}\right)^{-1+2i\nu}\psi^{(l,\nu)}(\rho)\,,
\end{empheq}
with
\begin{empheq}{alignat=7}
	\psi^{(l,\nu)}(\rho)&=
	\left(
	\begin{array}{c}
		{\displaystyle-\frac{\nu}{l+\frac{1}{2}}\left(\frac{\lambda(\nu)+m}{\lambda(\nu)-m}\right)^{\frac{1}{4}}\tanh\frac{\rho}{2}F\left(l+\frac{1}{2}-i\nu,1-i\nu;l+\frac{3}{2};\tanh^2\frac{\rho}{2}\right)}
		\\
		{\displaystyle i\left(\frac{\lambda(\nu)-m}{\lambda(\nu)+m}\right)^{\frac{1}{4}}F\left(l+\frac{1}{2}-i\nu,-i\nu;l+\frac{1}{2};\tanh^2\frac{\rho}{2}\right)}
	\end{array}
	\right)\,,
\end{empheq}
is regular at the origin. As before, the condition $\nu\in\mathds{R}$ is imposed by square-integrability. The solutions for the remaining three sectors can be obtained by simple operations, namely,
\begin{empheq}{alignat=7}\label{remsectors}
	l&\leq-\frac{1}{2}\,,
	&\quad
	\lambda(\nu)&>0
	&\qquad\longrightarrow\qquad
	&\left(i\sigma_1\right)h_{\pm}^{(-l,\nu)}(\rho)\Big|_{m\rightarrow-m}\,,
	\\\nonumber
	l&\geq\frac{1}{2}\,,
	&\quad
	\lambda(\nu)&<0
	&\qquad\longrightarrow\qquad
	&\left(i\sigma_2\right)h_{\pm}^{(l,\nu)}(\rho)\,,
	\\\nonumber
	l&\leq-\frac{1}{2}\,,
	&\quad
	\lambda(\nu)&<0
	&\qquad\longrightarrow\qquad
	&\left(i\sigma_3\right)h_{\pm}^{(-l,\nu)}(\rho)\Big|_{m\rightarrow-m}\,.
\end{empheq}
The normalization of the eigenspinors has been chosen so that they satisfy
\begin{empheq}{alignat=7}\label{eq: Wronskian fermions}
	h_-^{(l,\nu)}(\rho){h_+^{(l,\nu)}(\rho)}^T-h_+^{(l,\nu)}(\rho){h_-^{(l,\nu)}(\rho)}^T&=\frac{2i\sigma_2}{\sinh\rho}\,,
\end{empheq}
in all four sectors. This identity allow us to construct the Green's function
\begin{empheq}{alignat=7}\label{eq: Green's function fermions}
	G^{(l,\nu)}(\rho,\rho')&=\frac{i}{2}\sinh\rho'\left[h_-^{(l,\nu)}(\rho){h_+^{(l,\nu)}(\rho')}^T-h_+^{(l,\nu)}(\rho){h_-^{(l,\nu)}(\rho')}^T\right]\sigma_3\,\theta(\rho-\rho')\,,
	\\\nonumber
	&=\sinh\rho'\left[\phi^{\rm free}_{(l,\nu)}(\rho){h_+^{(l,\nu)}(\rho')}^T-h_+^{(l,\nu)}(\rho){\phi^{\rm free}_{(l,\nu)}(\rho')}^T\right]\sigma_3\,\theta(\rho-\rho')\,,
\end{empheq}
which satisfies
\begin{empheq}{alignat=7}
	\left(\mathcal{O}^{\textrm{free}}_l-\lambda(\nu)\right)G^{(l,\nu)}(\rho,\rho')&=-\delta(\rho,\rho')\,.
	\end{empheq}
Notice that
\begin{empheq}{alignat=7}
	G^{(l,\nu)}(\rho,\rho)&=-\frac{i}{2}\sigma_1\,,
\end{empheq}
as follows from the coincidence limit of the step function. Since we will need them shortly,  we present the asymptotic behavior of the solutions $h_+^{(l,i\nu)}(\rho)$ and $\phi^{\rm free}_{(l,i\nu)}(\rho)$ in the region where $(l+\frac12)\to\infty$ and $\nu=  \alpha (l+\frac12)$ with $0<\alpha<1$, 
\begin{empheq}{alignat=7}
h_+^{(l,i\nu)}(\rho) &\approx  \mathcal{F}(\rho) \, e^{-(l+\frac12) \eta(\rho)} \left(   \begin{array}{c} 1+\frac{1}{l+\frac12}\left(A(\rho)-\frac{i m}{2 \alpha}\right) \\
  \frac{-1+\sqrt{1+\alpha^2\sinh^2\rho}}{\alpha \sinh\rho}\left(1+\frac{1}{l+\frac12}\left(B(\rho)+\frac{i m}{2 \alpha}\right)\right)  \end{array}\right)\,,\label{psiasympt}\\
\phi^{\rm free}_{(l,i\nu)}(\rho)& \approx  \mathcal{G}(\rho) \, e^{(l+\frac12)\eta(\rho)} \left(   \begin{array}{c} 1+\frac{1}{l+\frac12}\left(C(\rho)-\frac{i m}{2 \alpha}\right) \\
 - \frac{1+\sqrt{1+\alpha^2\sinh^2\rho}}{\alpha \sinh\rho}\left(1+\frac{1}{l+\frac12}\left(D(\rho)+\frac{i m}{2 \alpha}\right)\right)  \end{array}\right) \, ,\label{jasympt} 
\end{empheq}
where $\eta(\rho)$ was defined in \eqref{etarho} and the rest of the functions involved satisfy the relations
\begin{empheq}{alignat=7}
\mathcal{F}(\rho)\mathcal{G}(\rho) &= \frac{i \alpha}{2\sqrt{1+\alpha^2\sinh^2\rho}}\,, &\qquad   
B(\rho)&= A(\rho) + \frac{1+\sqrt{1+\alpha^2\sinh^2\rho}}{2(1+\alpha^2\sinh^2\rho)}\\
C(\rho)&= -A(\rho)\,, &\qquad
D(\rho)&= -A(\rho)-\frac{-1+\sqrt{1+\alpha^2\sinh^2\rho}}{2(1+\alpha^2\sinh^2\rho)}\nn
\end{empheq}
As we will show below, the explicit forms of the functions $\mathcal{F}(\rho)$, $\mathcal{G}(\rho)$ and $A(\rho)$ do not play any role in the computation, so we do not present them here. Notice that we have included the first sub-dominant term.


\subsubsection{Regular Solution and Jost function}

We now consider the eigenvalue problem for the full operator \eqref{interacting fermion}.
The regular solution is constructed using the Lippmann-Schwinger equation, with the help of the free Green's function \eqref{eq: Green's function fermions},
\begin{empheq}{alignat=7}\label{eq: integral equation fermions}
	\phi_{(l,\nu)}(\rho)&=\phi^{\rm free}_{(l,\nu)}(\rho)+\int_0^{\rho}d\rho' G^{(l,\nu)}(\rho,\rho')\, \sigma_3\,U(\rho')\phi_{(l,\nu)}(\rho')\,.
\end{empheq}
Naively one would think that $\phi_{(l,\nu)}(\rho)\longrightarrow \phi^{\rm free}_{(l,\nu)}(\rho)$ as $\rho\rightarrow0$. However, a more careful analysis reveals that\footnote{Both $G^{(l,\nu)}(\rho,\rho)$ and $U(\rho)$ are finite at $\rho=0$, so the leading behavior is dictated by $\phi^{\rm free}_{(l,\nu)}(\rho)$. }
\begin{empheq}{alignat=7}
	\phi_{(l,\nu)}(\rho)&\approx \phi^{\rm free}_{(l,\nu)}(\rho)+G^{(l,\nu)}(\rho,\rho)\, \sigma_3 \, U(\rho)\int_0^{\rho}d\rho'\phi^{\rm free}_{(l,\nu)}(\rho')\nn
	\\
	&\approx\frac{i}{\Gamma\left(l+\frac{1}{2}\right)}\sqrt{\frac{\pi}{2}}\left|\frac{\Gamma\left(l+\frac{1}{2}- i\nu\right)}{\Gamma\left(\frac{1}{2}- i\nu\right)}\right|\left(\frac{\rho}{2}\right)^{l-\frac{1}{2}}\left(\frac{\lambda-m}{\lambda+m}\right)^{\frac{1}{4}}
	\left(
	\begin{array}{c}
		{\displaystyle i\frac{\lambda+m+V(0)-W(0)}{2l+1}\rho}
		\\
		{\displaystyle 1}
	\end{array}
	\right)\,.
\label{lpdnorm}
\end{empheq}
This is consistent with the behavior obtained by studying the differential equation near the origin. Accor\-dingly, for $l\leq -\frac12$ and $\lambda>0$, we have
\begin{empheq}{alignat=7}
	\phi_{(l,\nu)}(\rho)&\approx \frac{1}{\Gamma\left(|l|+\frac{1}{2}\right)}\sqrt{\frac{\pi}{2}}\left|\frac{\Gamma\left(|l|+\frac{1}{2} - i\nu\right)}{\Gamma\left(\frac{1}{2}- i\nu\right)}\right|\left(\frac{\rho}{2}\right)^{|l|-\frac{1}{2}}\left(\frac{\lambda+m}{\lambda-m}\right)^{\frac{1}{4}}
	\left(
	\begin{array}{c}
		{\displaystyle 1}
		\\
		{\displaystyle i\frac{\lambda-m-V(0)-W(0)}{2|l|+1}\rho}
	\end{array}
	\right)\,,\label{lpdnorm2}
	\end{empheq}
and similarly for the remaining two sectors. At any rate, the normalization of the regular solution is fixed by the normalization of the free eigenfunctions \eqref{eq: free eigenfunctions fermions}.

The Jost function can be extracted from the large $\rho$ behavior of the solution by means of its definition \eqref{jostdefinition}. 
A direct evaluation yields\footnote{As in the bosonic case, the effect of the shift in the Fourier mode due to the constant asymptotic value of the gauge field can be absorbed in the definition of the Jost function.}
\begin{empheq}{alignat=7}
	g_l(\nu)&=1+\int_0^{\infty}d\rho'\,\sinh\rho'h_+^{(l,\nu)}(\rho')^TU(\rho')\phi_{(l,\nu)}(\rho')\,.
\end{empheq}
As in the bosonic case, it will be sufficient to retain terms up to second order in the potential $U(\rho)$ so that
\begin{empheq}{alignat=7}
	\ln g_l(\nu) =&\int_0^{\infty}d\rho\,\sinh\rho\,h_+^{(l,\nu)}(\rho)^TU(\rho)\phi^{\rm free}_{(l,\nu)}(\rho)
	\label{perturbativejostfunction}\\
	&-\int_0^{\infty}d\rho\,\sinh\rho\,h_+^{(l,\nu)}(\rho)^TU(\rho)h_+^{(l,\nu)}(\rho)\int_0^{\rho}d\rho'\,\sinh\rho'\,\phi^{\rm free}_{(l,\nu)}(\rho')^TU(\rho')\phi^{\rm free}_{(l,\nu)}(\rho')+ \, O(U^3) \,.\nn
		\end{empheq}

We now need to continue the Jost function to imaginary radial momentum and extract its asymptotic behavior in the region $\left|l+\frac12\right|\to\infty$ and $\nu=  \alpha \left|l+\frac12\right|$ ($0<\alpha<1$). 
 In the sector of positive $l$ and positive $\lambda$ we can make use of the asymptotic expansions presented above.  The calculation proceeds much like the bosonic case with the proviso that the eigenfunctions have spinorial structure. However, the fermionic potential is $l$-independent and now subleading orders in \eqref{psiasympt}-\eqref{jasympt} do contribute. Again resorting to a saddle point approximation we find
\begin{align}
\ln g^{+}_l(i\nu)& = \frac{i\alpha}{2}\int d\rho  \frac{\sinh\rho  \left(U^{(0)}_{h\phi}+\frac{1}{l+\frac12 } U^{(1)}_{h\phi}\right)}{\sqrt{1+\alpha^2\sinh^2\rho}}
 + \frac{\alpha^2}{4(2l+1)}\int_0^\infty d\rho\, \frac{\sinh^3\rho \, U_{hh}\, \,U_{\phi\phi}}{(1+\alpha^2\sinh^2\rho)^\frac32} +  O\left(l^{-2}\right)\,,
\end{align} 
where
\begin{align}
U^{(0)}_{h\phi}&= (U_{11}-U_{22})-\frac{1}{\alpha\sinh\rho} (U_{12}+U_{21})-\frac{\sqrt{1+\alpha^2\sinh^2\rho}}{\alpha\sinh\rho}(U_{12}-U_{21})\,,\\
U^{(1)}_{h\phi}&= -\frac{i m}{\alpha}(U_{11}+U_{22})-\frac{U_{22}}{1+\alpha^2\sinh^2\rho}+\frac{\alpha\sinh\rho}{2(1+\alpha^2\sinh^2\rho)}(U_{12}+U_{21})\,,\nn\\
U_{hh}&= U_{11} + U_{22}\left(\frac{-1+\sqrt{1+\alpha^2\sinh^2\rho}}{\alpha\sinh\rho}\right)^2 + \frac{-1+\sqrt{1+\alpha^2\sinh^2\rho}}{\alpha\sinh\rho}(U_{12}+U_{21})\,,\nn\\
U_{\phi\phi}&= U_{11} + U_{22}\left(\frac{1+\sqrt{1+\alpha^2\sinh^2\rho}}{\alpha\sinh\rho}\right)^2 - \frac{1+\sqrt{1+\alpha^2\sinh^2\rho}}{\alpha\sinh\rho}(U_{12}+U_{21})\,.\nn
\end{align}
As was previously mentioned, these expressions are independent of the function $A(\rho)$ appearing in the asymptotic expansions of $h_+^{(l,\nu)}(\rho)$ and $\phi^{\rm free}_{(l,\nu)}(\rho)$.

The remaining three sectors of solutions are obtained by performing the operations \eqref{remsectors}, which amount to the substitutions $U\to(i \sigma_i)^T U (i\sigma_i)$ and $m\to\pm m$ in the above formul\ae. After summing over all four sectors and discarding a $\nu$-independent term we identify the potentially divergent part as
\begin{align}
\ln \mathfrak{g}^{\rm asym}_l(i\nu)+\ln \mathfrak{g}^{\rm asym}_{-l}(i\nu)&\equiv
\frac{2}{l+\frac12}\int_0^\infty d\rho \sinh\rho \frac{ \left(U_{11}+m\right)\left(U_{22}+m\right)-m^2}{\sqrt{1+\alpha^2\sinh^2\rho}} \nn\\
& + \frac{\alpha^2}{2(l+\frac12)} \int_0^\infty d\rho \sinh^3\rho \frac{\left(U_{11}-U_{22}\right)^2-\left(U_{12}+U_{21}\right)^2}{\left(1+\alpha^2\sinh^2\rho\right)^{\frac32}}\,,
\label{gasymtotal}
\end{align}
where we made use of the definition below \eqref{eq: zeta function fermions Jost}. Note that 
\be 
\lim_{\nu\to\infty}\left(\ln \mathfrak{g}^{\rm asym}_l(i\nu)+\ln \mathfrak{g}^{\rm asym}_{-l}(i\nu)\right)=0\,.
\label{gAsInf}
\ee
On the other hand, a similar calculation but in the limit of large $\nu$ and fixed $l$ yields
\be 
\lim_{\nu\to\infty}\left(\ln \mathfrak{g}_l\left(i\nu\right)+\ln \mathfrak{g}_{-l}\left(i\nu\right)\right) = 2 i \int d\rho \left(U_{21}-U_{12}\right)\,,
\label{ginfty}
\ee 
which is non-vanishing. This is an effect of the normalization \eqref{lpdnorm}.


\subsubsection{Analytic continuation }
The analytic continuation of \eqref{eq: zeta function fermions Jost} proceeds much in the same way as for bosons. We split the sum over Fourier modes as
\begin{empheq}{alignat=7}
	\hat{\zeta}_{\mathcal{O}^2}(s)&=\hat{\zeta}_f(s)+\hat{\zeta}_d(s)\,,
\end{empheq}
where
\begin{empheq}{alignat=7}
\hat{\zeta}_f(s)&=\frac{\sin\pi s}{\pi}\sum_{l=\frac12}^{\infty}\int_{\nu_0}^{\infty}d\nu\left(\nu^2-\nu_0^2\right)^{-s}\partial_{\nu}\left(\ln \mathfrak g_l(i\nu)+\ln \mathfrak g_{-l}(i\nu)-\ln \mathfrak g^{\textrm{asym}}_l(i\nu)-\ln\mathfrak  g^{\rm asym}_{-l}(i\nu)\right)\,,
	\\
	\hat{\zeta}_d(s)&=\frac{\sin\pi s}{\pi}\sum_{l=\frac12}^{\infty}\int_{\nu_0}^{\infty}d\nu\left(\nu^2-\nu_0^2\right)^{-s}\partial_{\nu}\left(\ln\mathfrak g^{\textrm{asym}}_l(i\nu)+\ln\mathfrak g^{\rm asym}_{-l}(i\nu)\right)\,.
\end{empheq}
The series in $\zeta_f(s)$ is now convergent at $s=0$ and we find
\begin{empheq}{alignat=7}
	\hat{\zeta}_f(0)&=0\,,
	\\\label{zetafinite}
	\hat{\zeta}'_f(0)&=-\sum_{l=\frac{1}{2}}^{\infty}\left(\ln\mathfrak{g}_l(i\nu_0)+\ln\mathfrak{g}_{-l}(i\nu_0)-\ln\mathfrak{g}^{\textrm{asym}}_l(i\nu_0)-\ln\mathfrak{g}^{\textrm{asym}}_{-l}(i\nu_0)-2i\left(U_{12}-U_{21}\right)\right)\,,
\end{empheq}
Were it not for the last term, coming from \eqref{ginfty}, the sum over Fourier modes would suffer from a linear divergence. In turn, to compute $\zeta_d(s)$ we make use of the asymptotic form of the Jost function given in \eqref{gasymtotal} and the results \eqref{intform1}-\eqref{intform2} to perform the momentum integrals, thus obtaining
\begin{align}
\hat{\zeta}_d(s)=& -\frac{2\Gamma\left(s+\frac12\right)\Gamma\left(1-s\right)}{\Gamma\left(\frac12 \right)}
\int_0^\infty  d\rho\,(\sinh\rho)^{2s+1}\left(\left(U_{11}+m\right)\left(U_{22}+m\right)-m^2\right)R_1(s,\rho)\\
& -\frac{\Gamma\left(s+\frac12\right)\Gamma\left(1-s\right)}{2\Gamma\left(\frac12\right)}\int_0^\infty  d\rho\,(\sinh\rho)^{2s+1}\left(\left(U_{11}-U_{22}\right)^2-\left(U_{12}+U_{21}\right)^2\right)R_{2}(s,\rho)\,.\nn
\end{align}
The sums $R_1(s,\rho)$ and $R_2(s,\rho)$ become equal to \eqref{R1} and \eqref{R2}, respectively, after shifting $l\rightarrow l-\frac{1}{2}\in\mathds{N}^+$ and using $\nu_0=m$. The shift is a legal operation since we have not set $s=0$ yet and the sums are convergent. Surely, the continuation of $R_1(s,\rho)$ and $R_2(s,\rho)$ is the same as before. Hence we arrive at
\begin{empheq}{alignat=7}
	\zeta_d(0)=&-\int_0^{\infty}d\rho\,\sinh\rho\left(\left(U_{11}+m\right)\left(U_{22}+m\right)-m^2\right)\nn\\
	\hat{\zeta}_d'(0)=&-2\int_0^{\infty}d\rho\,\sinh\rho\left(\ln\left(\frac{\sinh\rho}{2}\right)+\gamma\right)\,\left(\left(U_{11}+m\right)\left(U_{22}+m\right)-m^2\right)
	\\
	& +\frac{1}{2}\int_0^{\infty}d\rho\,\sinh\rho\,\left(\left(U_{11}-U_{22}\right)^2-\left(U_{12}+U_{21}\right)^2\right)
	\nn\\
	& -\sum_{l=\frac{1}{2}}^{\infty}\left(\ln\mathfrak  g^{\textrm{asym}}_l\left(i\nu_0\right)+\ln\mathfrak  g^{\textrm{asym}}_{-l}\left(i\nu_0\right)-\frac{2}{l+\frac{1}{2}}\int_0^{\infty}d\rho\,\rho\,\left(\left(U_{11}+m\right)\left(U_{22}+m\right)-m^2\right)\right)\,,\nn
\end{empheq}
where we have used the expression \eqref{gasymtotal} to recognize $\ln\mathfrak  g^{\textrm{asym}}_l\left(i\nu_0\right)+\ln \mathfrak g^{\textrm{asym}}_{-l}\left(i\nu_0\right)$. Collecting all the pieces we obtain
\begin{empheq}{alignat=7}
\hat{\zeta}_{\mathcal{O}^2}(0)=&-\int_0^{\infty}d\rho\,\sinh\rho \left(\left(U_{11}+m\right)\left(U_{22}+m\right)-m^2\right)\nn\\
\hat{\zeta}'_{\mathcal{O}^2}(0) =& -2\sum_{l=\frac12}^{\infty}\left(\ln g^+_l\left(i\nu_0\right)+\ln g^+_{-l}\left(i\nu_0\right) - i \int d\rho \left(U_{21}-U_{12}\right) +\frac{1}{l+\frac12}\zeta_{\mathcal{O}^2}(0)\right)+2\gamma\hat{\zeta}_{\mathcal{O}^2}(0) \nn\\
& -2\int_0^{\infty}d\rho\,\sinh\rho\ln\left(\frac{\sinh\rho}{2}\right)\left(\left(U_{11}+m\right)\left(U_{22}+m\right)-m^2\right)
	\\
	& +\frac{1}{2}\int_0^{\infty}d\rho\,\sinh\rho\,\left(\left(U_{11}-U_{22}\right)^2-\left(U_{12}+U_{21}\right)^2\right)\nn \, ,
\end{empheq}
where we have made explicit that since $\lambda(i\nu_0)=0$, the Jost functions $g^+_l(i\nu_0)$ and $g^-_l(i\nu_0)$ coincide. Finally, through \eqref{Jostsolution}, \eqref{eq: zetap 1d fermions} and \eqref{ginfty} we identify
\begin{empheq}{alignat=7}
	\hat{\zeta}'_{\mathcal{O}_l^2}(0)&=-2\ln g^+_l(i\nu_0)-i\int_0^{\infty}d\rho\left(U_{12}-U_{21}\right)\,.
\end{empheq}
Writing the potential components in terms of the background fields and recalling that $\hat{\zeta}_{\mathcal{O}}(s)=\frac{1}{2}\hat{\zeta}_{\mathcal{O}^2}(s)$ we arrive at our main result \eqref{eq: main result fermions} for the determinant of a fermionic operator.


 \section{Conclusions}\label{sec: conclusions}

 In this manuscript we have explicitly computed the determinants for a general class of circularly-symmetric bosonic and fermionic  operators in $AdS_2$ and spaces that are conformally $AdS_2$.  In this context there are a number of options depending on the regularization technique used. Some widely used regularization techniques are not explicitly diffeormophism invariant.  Our main result is to have obtained  answers that are completely aligned with the zeta-function regularization method.  Consequently, and importantly, we now have diffeormphic-invariant expressions for such determinants.

Our driving motivation has been to enlarge the arsenal of tools required to push the  AdS/CFT correspondence into its precision regime.  An important limitation of our computation is that it exploits, in a crucial manner, the angular symmetry of the problem. Namely, we are able to turn the problem into  effectively a one-dimensional one due to the symmetry. There are many problems in this class, some we have mentioned but others are less obvious such as the one-loop correction to the anti-parallel lines. It would be interesting, however, to have a better understanding of the form of the determinant independently of the symmetries and ultimately a computational approach that is intrinsically two-dimensional.  The drive to less symmetric situations is not merely an academic goal. There are examples which are under control from the localization point of view but where the symmetry is not preserved \cite{Giombi:2012ep}. More general methods are still needed and it would be valuable to develop them.

Precision holography has largely focused on the results provided by supersymmetric localization. It would be great to connect with the efforts developed in the context of integrability \cite{Gromov:2007aq},\cite{Beisert:2010jr}.  Integrability provides a wide field to explore from the point of semi-classical gravity computations. Ultimately, one would hope to tackle questions with less or no supersymmetry and where integrability does not play a role.  We also expect that our methods will find use in other problems possibly related to one-loop gravity computations in the context of corrections to  black hole entropy, as determinants in  $AdS_2$ have already been found in many works starting with \cite{Sen:2011ba} and its sequels.


\section*{Acknowledgments}
LPZ, VR and GAS thank ICTP for providing  hospitality at various stages. 
AF was supported by Fondecyt \# 1160282.  LPZ and VR are partially supported by the US Department of Energy under Grant No. DE-SC0017808 –{\it  Topics in the AdS/CFT Correspondence: Precision tests with Wilson loops, quantum black holes and dualities.} GAS and JAD are supported by CONICET and grants PICT
2012-0417, PIP0595/13, X648 UNLP, PIP 0681, PIP 2017-1109 and PI {\it B\'usqueda de nueva F\'isica}.


\appendix

\section{Weyl Anomaly}
\label{app: Weyl Anomaly}

In two dimensions, for an operator of the form
\begin{empheq}{alignat=7}
	\mathcal{O}_M&=M^{-1}\mathcal{O}\,,
	&\qquad
	\mathcal{O}&=-g^{\mu\nu}D_{\mu}D_{\nu}+X\,,
\end{empheq}
the dependence of $\textrm{det}\,\mathcal{O}_M$ on $M$ is determined by \cite{Drukker:2000ep,Vassilevich:2003xt,Schwarz:1979ae}
\begin{empheq}{alignat=7}
	\delta_M\left(\ln\det\mathcal{O}_M\right)&=-a_2\left(\delta\ln M|\mathcal{O}_M\right)\,,
\end{empheq}
where $a_2$ is the Seeley coefficient 
\begin{empheq}{alignat=7}
	a_2(F|\mathcal{O}_M)&=\frac{1}{4\pi}\textrm{Tr}\left[\int_{\cal M} d^2\sigma\sqrt{g}\,F\,b_2(\mathcal{O}_M)+\int_{\partial\cal M}ds\sqrt{\gamma}\left(F\,c_2(\mathcal{O}_M)\mp\frac12\partial_n F\right)\right]\,,\\
	b_2(\mathcal{O}_M)&= -X+\frac{1}{6}R-\frac{1}{6}\nabla^2\ln M \,,\qquad c_2(\mathcal{O}_M)=\frac13\left(K-\frac12\partial_n\ln M\right)\,,
\end{empheq}
and the trace is taken over all degrees of freedom. For $AdS_2$ the unit normal vector and the extrinsic curvature are given by $	n=\partial_\rho$ and $ K= g^{\mu\nu}\nabla_\mu n_\nu =\coth\rho$.  Integrating this relation yields
\begin{empheq}{alignat=7}
	\ln\left(\frac{\det\mathcal{O}_{M}}{\det\mathcal{O}}\right)&=\frac{1}{4\pi}\int d^2\sigma\sqrt{g}\ln M\,\textrm{Tr}\left(X-\frac{1}{6}R+\frac{1}{12}\nabla^2\ln M\right)\,.
\end{empheq}
Here we have discarded boundary terms, which is justified as long as the conformal factor is everywhere smooth with $M\rightarrow1$ sufficiently fast as $\rho\rightarrow\infty$. This is all that is needed for the scalar case. 

The treatment of fermionic fluctuations is similar, except that the anomaly argument only works for second order operators. So, given instead
\begin{empheq}{alignat=7}
	\mathcal{O}_M&=M^{-\frac{1}{2}}\mathcal{O}\,,
	&\qquad
	\mathcal{O}&=-i\slashed{D}+Y\,.
\end{empheq}
we must relate the determinants of $\mathcal{O}_M^2$ and $\mathcal{O}^2$. Directly squaring leads to
\begin{empheq}{alignat=7}
	\mathcal{O}^2_M&=M^{-1}\mathcal{O}'\,,
	&\qquad
	\mathcal{O}'&=-g^{\mu\nu}D_{\mu}'D_{\nu}'+X'\,,
\end{empheq}
where
\begin{empheq}{alignat=7}
	D'_{\mu}&=D_{\mu}+\frac{i}{2}\theta_{\mu}\,,
	&\qquad	
	\theta_{\mu}&=\Gamma_{\mu}Y+Y\Gamma_{\mu}+\frac{i\slashed{\partial}M}{2M}\Gamma_{\mu}\,.
\end{empheq}
and
\begin{empheq}{alignat=7}
	X'&=-\frac{1}{4}\left(\Gamma^{\mu}Y\Gamma_{\mu}Y+Y\Gamma^{\mu}Y\Gamma_{\mu}+\Gamma^{\mu}Y^2\Gamma_{\mu}-2Y^2+\Gamma_{\mu}Y\frac{i\slashed{\partial}M}{2M}\Gamma^{\mu}-\frac{i\slashed{\partial}M}{M}Y+\frac{i\slashed{\partial}M}{2M}\Gamma^{\mu}Y\Gamma_{\mu}\right)
	\\
	&+\frac{i}{2}\left(-\Gamma^{\mu}D_{\mu}Y+D_{\mu}Y\Gamma^{\mu}+\frac{i}{2}\nabla^2\ln M\right)+\frac{1}{4}R-iq\slashed{\mathcal{F}}\,.
\end{empheq}
The corresponding Seeley coefficient reads
\begin{empheq}{alignat=7}
	\textrm{Tr}\, b_2(\mathcal{O}_M^2)&=\textrm{Tr}\left(-X'+\frac{1}{6}R-\frac{1}{6}\nabla^2\ln M\right)&&=\textrm{Tr}\left(\frac{1}{2}\Gamma^{\mu}Y\Gamma_{\mu}Y-\frac{1}{12}R+\frac{1}{12}\nabla^2\ln M\right)\,.
\end{empheq}
Integrating the anomaly equation yields
\begin{empheq}{alignat=7}
	\ln\left(\frac{\det\mathcal{O}^2_{M}}{\det\mathcal{O}^2}\right)&=\frac{1}{4\pi}\int d^2\sigma\sqrt{g}\,\ln M\,\textrm{Tr}\left(-\frac{1}{2}\Gamma^{\mu}Y\Gamma_{\mu}Y+\frac{1}{12}R-\frac{1}{24}\nabla^2\ln M\right)\,.
\end{empheq}

\bibliographystyle{JHEP}
\bibliography{Bib}
\end{document}